\newif\ifdraft

\draftfalse
\InputIfFileExists{localflags}{}

\ifdraft
\fi

\PassOptionsToPackage{table,dvipsnames}{xcolor}
\documentclass[sigconf,nonacm=true,anonymous=false,natbib=false,balance=true,pbalance=false]{acmart}

\newif\iffullversion
\fullversiontrue

\usepackage{versions}
\includeversion{full}
\excludeversion{conf}

\usepackage{microtype}
\usepackage{graphicx}
\usepackage[dvipsnames]{xcolor}
\input{macros/macros-listings-tamarin.sty}
\usepackage{amsthm}
\usepackage{mathtools}
\usepackage{bm}
\usepackage{csquotes}
\usepackage{booktabs}
\usepackage{array}
\usepackage{tabularx}
\usepackage{multirow}
\usepackage{siunitx}
\usepackage[inline]{enumitem}
\usepackage{listings}
\AtBeginDocument{%
  \lstset{
    numbers=none,
    frame=none,
    xleftmargin=1.5em
  }%
}
\usepackage{pifont}
\usepackage{xr}
\usepackage[nameinlink,capitalise,noabbrev]{cleveref}
\usepackage[font=footnotesize,labelsep=period]{caption}
\usepackage{subcaption}

\usepackage[
  backend=biber,
  style=acmnumeric,
  doi=false,
  url=true,
  isbn=false,
  eprint=false
]{biblatex}

\DeclareFieldFormat{titlecase}{#1}

\DeclareFieldFormat{pages}{\mbox{#1}}

\newsavebox{\biburlbox}
\DeclareFieldFormat{url}{%
  \sbox{\biburlbox}{\url{#1}}%
  \ifdim\wd\biburlbox<0.4\linewidth
    \usebox{\biburlbox}%
  \else
    \url{#1}%
  \fi}

\renewbibmacro*{swurl+urldate}{%
  \iffieldundef{urlyear}{}{\usebibmacro{urldate}\addspace}%
  \ifhyperref
    {\href{\strfield{url}}{\nolinkurl{\strfield{url}}}}%
    {\edef\temp{\noexpand\nolinkurl{\strfield{url}}}\temp}}

\DeclareSourcemap{
  \maps[datatype=bibtex]{
    \map{
      \pernottype{online}
      \pernottype{software}
      \step[fieldsource=entrykey, notmatch=\regexp{^(cheval2022|full)$}, final]
      \step[fieldset=url, null]
      \step[fieldset=urldate, null]
    }
    \map{
      \step[fieldset=isbn, null]
      \step[fieldset=issn, null]
      \step[fieldset=eprint, null]
      \step[fieldset=eprintclass, null]
      \step[fieldset=eprinttype, null]
      \step[fieldset=editor, null]
      \step[fieldset=organization, null]
    }
    \map{
      \step[fieldsource=entrykey, notmatch=\regexp{^said2026$}, final]
      \step[fieldset=doi, null]
    }
    \map{
      \pertype{book}
      \step[fieldset=series, null]
    }
    \map{
      \pertype{inproceedings}
      \pertype{incollection}
      \step[fieldset=series, null]
      \step[fieldset=volume, null]
      \step[fieldset=number, null]
      \step[fieldset=month, null]
      \step[fieldset=address, null]
      \step[fieldset=location, null]
      \step[fieldset=eventtitle, null]
      \step[fieldsource=date, match=\regexp{^(\d{4})-.*$}, replace={$1}]
      \step[fieldsource=booktitle, match=\regexp{,\s+Part\s+\{?[IVX]+\}?\s*$}, replace={}]
    }
  }
}

\AtBeginBibliography{\setlength{\emergencystretch}{2em}}
\renewbibmacro*{finentry}{\finentry\par}
\AtEveryBibitem{%
  \ifboolexpr{
    test {\ifentrytype{misc}} or test {\ifentrytype{thesis}}
  }{\renewbibmacro*{date}{}}{}%
  \iffieldequalstr{entrykey}{qi2024}
    {\DeclareListFormat{institution}{\mbox{#1}}}{}%
  \ifboolexpr{
    test {\iffieldequalstr{entrykey}{hashimoto2025}} or
    test {\iffieldequalstr{entrykey}{libcrux-mlkem}} or
    test {\iffieldequalstr{entrykey}{goldberg2025}} or
    test {\iffieldequalstr{entrykey}{curve25519-dalek}}
  }{\looseness=-1\relax}{}%
  \ifboolexpr{
    test {\iffieldequalstr{entrykey}{cade2015}} or
    test {\iffieldequalstr{entrykey}{cohn-gordon2016}} or
    test {\iffieldequalstr{entrykey}{whatsapp-web}}
  }{%
    \setlength{\rightskip}{0pt plus 2em}%
    \setlength{\parfillskip}{0pt plus 0.5\linewidth}%
    \Urlmuskip=0mu\relax
    \finalhyphendemerits=100000\relax
  }{}%
  \ifboolexpr{
    test {\iffieldequalstr{entrykey}{marlinspike2016}} or
    test {\iffieldequalstr{entrykey}{codeverify}}
  }{%
    \setlength{\parfillskip}{0pt plus 0.5\linewidth}%
    \useOriginalUrlSetting
    \Urlmuskip=0mu\relax
  }{}%
}

\makeatletter
\def\@formatdoi#1{\allowbreak\mbox{\url{https://doi.org/#1}}}
\patchcmd{\@mkbibcitation}{\bgroup}{\bgroup\emergencystretch=2em\relax}
  {}{\PackageError{paper}{Could not adjust ACM reference line breaking}{}}
\makeatother
\iffullversion
\fi

\newcommand{\fullversionurl}{https://arxiv.org/pdf/2609.11882}
\newcommand{\fullversionlocation}[1]{%
  \href{\fullversionurl\#nameddest=\getrefbykeydefault{F-#1}{anchor}{}}{\Cref*{F-#1}}%
}

\newcommand{\xmark}{\textcolor{Red}{\ding{55}}}
\newcommand{\cmark}{\textcolor{Green}{\ding{51}}}

\newcommand{\term}[1]{\ensuremath{\mathit{#1}}}

\newcommand{\libsignal}{\ensuremath{\mathtt{libsignal}}\xspace}
\newcommand{\asignal}{$\mathcal{A}_\textrm{Signal}$\xspace}
\newcommand{\awhatsapp}{$\mathcal{A}_\textrm{WhatsApp}$\xspace}
\newcommand{\apcs}{$\mathcal{A}_\textrm{PCS}$\xspace}

\newcommand{\z}[1]{\ensuremath{\mathit{#1}}}
\newcommand{\whatsapp}{WhatsApp\xspace}
\newcommand{\signal}{Signal\xspace}

\newcommand{\var}[1]{\z{#1}}
\newcommand{\tvar}[1]{\z{#1}}

\definecolor{bluegray}{rgb}{0.4, 0.6, 0.8}
\newcommand{\commentRBlue}[1]{\textcolor{bluegray}{\text{\emph{ // #1}}}}

\newcommand{\aenc}{\mathit{aenc}}
\newcommand{\adec}{\mathit{adec}}

\newcommand{\LTK}{\mathit{LTK}}
\newcommand{\SPK}{\mathit{SPK}}
\newcommand{\OPK}{\mathit{OPK}}
\newcommand{\opk}{\mathit{opk}}
\newcommand{\sk}{\mathit{sk}}
\newcommand{\SK}{\mathit{SK}}
\newcommand{\mlkemsk}{\mathit{mlkemsk}}
\newcommand{\mlkemPK}{\mathit{mlkemPK}}
\newcommand{\kybersk}{\mlkemsk}
\newcommand{\kyberPK}{\mlkemPK}
\newcommand{\ltka}{\z{\LTK_{A}}}
\newcommand{\ltkb}{\z{\LTK_{B}}}
\newcommand{\spkb}{\z{\SPK_{B}}}
\newcommand{\opkb}{\z{\OPK_{B}}}
\newcommand{\mlkemb}{\z{mlkemPK_{B}}}
\newcommand{\kyberb}{\mlkemb}

\iffullversion\else
\fi

\theoremstyle{definition}

\usepackage{macros/macros-meta}
\usepackage{macros/macros-msr}
\usepackage{pgfplots}
\usepackage{graphicx}

\pgfplotsset{compat=1.18}

\title[From Specs to Apps: Verifying and Monitoring Models of Signal and WhatsApp]{From Specs to Apps: Verifying and Monitoring Models\texorpdfstring{\\}{ }of Signal and WhatsApp}
\iffullversion
\subtitle{Full Version\footnotemark}
\else
\copyrightyear{2026}
\acmYear{2026}
\setcopyright{cc}
\setcctype{by}
\acmConference[CCS '26]{Proceedings of the 2026 ACM SIGSAC Conference on Computer and Communications Security}{November 15--19, 2026}{The Hague, Netherlands}
\acmBooktitle{Proceedings of the 2026 ACM SIGSAC Conference on Computer and Communications Security (CCS '26), November 15--19, 2026, The Hague, Netherlands}
\acmDOI{10.1145/3830454.3846691}
\acmISBN{979-8-4007-2871-6/2026/11}

\fi

\author{Moustafa Said}
\orcid{0009-0003-8874-0424}
\email{moustafa.said@cispa.de}
\affiliation{%
	\institution{CISPA Helmholtz Center for Information Security}
	\city{Saarbrücken}
	\country{Germany}
}

\author{Aurora Naska}
\orcid{0009-0005-9528-656X}
\email{aurora.naska@cispa.de}
\affiliation{%
	\institution{CISPA Helmholtz Center for Information Security}
	\city{Saarbrücken}
	\country{Germany}
}

\author{Kevin Morio}
\orcid{0000-0002-0220-3448}
\email{kevin.morio@cispa.de}
\affiliation{%
	\institution{CISPA Helmholtz Center for Information Security}
	\city{Saarbrücken}
	\country{Germany}
}

\author{Robert Künnemann}
\orcid{0000-0003-0822-9283}
\email{robert.kuennemann@cispa.de}
\affiliation{%
	\institution{CISPA Helmholtz Center for Information Security}
	\city{Saarbrücken}
	\country{Germany}
}

\iffullversion\else
\fi

\hypersetup{
  pdftitle={From Specs to Apps: Verifying and Monitoring Models of Signal and WhatsApp}
}

\iffullversion
\else
\fi
\begin{document}

\begin{abstract}
  The Signal protocol is a prominent messaging protocol that secures communication for billions of users.
It powers WhatsApp, the most widely used messaging application worldwide,
and the Signal app, popular among privacy-conscious users.
Extensive research in the computational and Dolev-Yao settings provides strong formal security guarantees for the protocol itself.
However, a gap remains between the guarantees of the protocol specification and the implementation's actual behavior at runtime.

In this work, we bridge this gap by applying SpecMon, a recently
proposed runtime monitor, to check whether observed executions conform
to formal protocol models.
To this end, we instrument two applications (WhatsApp Web and Signal Desktop) to capture their interactions with the network and the cryptographic components. Using this instrumentation, we develop two multiset-rewrite models that are compatible with Tamarin, thus enabling verification.
We derive the first model of WhatsApp Web's implementation of the Signal protocol and the most detailed model to date of Signal's original protocol. Monitoring establishes that observed executions conform to these models, relative to the trusted event extraction and the symbolic abstraction. For
the core components of the Signal protocol, we verify
authentication and secrecy properties.
Finally, monitoring 
reveals previously undocumented differences between the original \libsignal{} library and WhatsApp's fork.

We evaluate our methodology and demonstrate its reproducibility. Developing the WhatsApp Web model, instrumenting the app, adding fuzzing, and running the experiments took three person-weeks.
We also demonstrate efficient monitoring of real-world applications and detection of deliberately injected security faults, with low overhead in our measured setting.

\end{abstract}

\maketitle

\iffullversion
  \let\savedthefootnote\thefootnote
  \renewcommand{\thefootnote}{}
  \footnotetext{* This is an extended version of~\cite{said2026}.}
  \let\thefootnote\savedthefootnote
\fi

\section{Introduction}
\label{sec:introduction}

The Signal protocol is a prominent messaging protocol that secures the daily communications of billions of users worldwide.
This open-source protocol~\cite{libsignal} serves as the backbone of a multitude of modern messaging apps, including Signal~\cite{signal-org}, WhatsApp~\cite{wa}, Facebook Messenger~\cite{facebook-wp}, and Google Messages~\cite{google-wp}. 


Due to its prevalence, there has been extensive research~\cite{beguinetFormalVerificationPostquantum2024,cohn-gordonFormalSecurityAnalysis,cremers2023a,
froschHowSecureTextSecure2014,kobeissi2017,
marlinspikeSesameAlgorithmSession2017,
vandamAnalysingSignalProtocol,
DBLP:conf/eurocrypt/AlwenCD19, canetti2022universally, 
DBLP:conf/crypto/BienstockFGMR22,
hashimoto2025bundled, fiedler2025security, linker2025looping} to formally analyze the protocol. 
In particular, the use of state-of-the-art automated verification tools
like Tamarin~\cite{meier2013} and ProVerif~\cite{blanchet2016modeling} 
has enabled analysis of some of the more complex security guarantees of the protocol~\cite{bhargavanFormalVerificationPQXDH, cremers2023a, linker2025looping}.
However, keeping verification tractable often requires abstracting protocol details
or focusing on a particular component instead of the entire system.

This leaves a verification gap: guarantees proved on a formal model do not transfer to the actual implementation of the protocol.
The implementation may deviate from the model in various ways, e.g., by omitting security-relevant steps, by using different cryptographic primitives, or by introducing bugs.
To further complicate matters, although messaging apps use an open-source protocol, they are typically closed-source, making it even more difficult to build trust that the theoretical guarantees 
of the underlying protocol transfer to the application. 
Facebook (now Meta), the current owner of WhatsApp, was named in the NSA's PRISM disclosures~\cite{greenwald-prism}, which contributed to long-standing concerns among security-conscious users about WhatsApp's trustworthiness.
%
Nevertheless,
both Signal~\cite{goldbergTrumpAdministrationAccidentally2025,
michaelPhotosRevealTrump2025}
and
WhatsApp~\cite{ban}
are still used for highly classified communication despite government restrictions.

It is hence fundamental to ensure that the implementation conforms to the verified model.
Techniques for doing so include
code verification~\cite{sprenger2020},
verified compilers (cv2ocaml~\cite{cade2015}, cv2fstar~\cite{lipp2022}, and the compiler of~\cite{almeida2013}),
type checking (F*~\cite{swamy2011}),
model extraction~\cite{aizatulin2012, kobeissi2017, nasrabadi2023},
(automated) theorem proving~\cite{jurjens2008}, and
refinement from specifications~\cite{bhargavan2010,polikarpova2012,arquint2022}.
In contrast to these static techniques, which either require
substantial proof effort or apply only to limited implementations,
\emph{dynamic verification} (runtime monitoring) provides a flexible,
black-box view of the program.
SpecMon~\cite{specmon}, a recently proposed runtime monitoring engine, addresses this problem by checking compliance with a formal model during the execution of a protocol implementation. Within the bounds of the Dolev-Yao (DY) model and the monitored scope, guarantees of the formal model transfer to accepted implementation traces.

In this work, we establish formal
models that allow SpecMon to monitor secure messaging applications. This entails two key
tasks. First, the implementation needs to be \emph{instrumented},
i.e., cryptographic components and network interfaces are
identified and their inputs and outputs are exposed to the monitor.
Second, the model needs specification-level information
that verification models sometimes omit. In theory, this only requires a description of the concrete
on-the-wire format (SpecMon provides a language to express this that is
backward compatible with Tamarin). In practice, formal models often
omit potentially critical details for the sake of verification. To
capture the desired behaviors of the implementation, more detail can
be necessary. 

Our work targets two prominent Signal protocol implementations: WhatsApp Web, version \texttt{2.3000.1029798056} (November 2025), and Signal Desktop, version \texttt{8.4.0-alpha.1} (April 2026). Our Signal Desktop experiments use \libsignal{} \texttt{v0.88.1}.

In our first case study, we monitor Signal Desktop~\cite{signal-d}, an open-source client,
to detect deviations during application runtime.
Starting from an initial model of 
Signal's session-handling layer, Extended Triple Diffie-Hellman (X3DH), and Double Ratchet (DR)~\cite{cremers2023a, albert},
we enhance the model with the post-quantum key exchange (PQXDH)~\cite{pqxdh}, now supported by Signal, Sealed Sender~\cite{sealed-sender-doc}, and other details omitted in previous work, such as cipher-key derivations.
This leads to the most detailed model for Signal to date, which is validated against the actual implementation.

In our second case study, we show that this methodology also applies to closed-source implementations.
%
%
We monitor a WhatsApp client~\cite{wa} using Chrome DevTools~\cite{devtools}, instrumenting WhatsApp's modified version of \libsignal{}.
Starting from traces generated by the application and no previous model,
we extract the first formal model of WhatsApp's implementation of the Signal protocol.

For both models, we use Tamarin to prove authentication and secrecy of the initial root key under the threat models described in \cref{sec:verification}. For Signal, the analysis also considers an attacker that can break the Diffie-Hellman (DH) assumption. The authentication guarantee does not cover a DH break before the responder completes the handshake. We also instantiate an impossibility result~\cite{DBLP:conf/sp/CremersMN25} against post-compromise security (PCS, the guarantee that a conversation's security can be restored after its secrets are compromised) in our setting.
Additionally, we point out undocumented differences between Signal's and WhatsApp's DR implementations.

Our work demonstrates the practical value of monitoring real-world applications. 
First, our methodology enables fast and efficient model extraction of complex protocols: developing the WhatsApp Web model, instrumenting the app, adding fuzzing, and running the experiments took three person-weeks.
Second, and more significantly, maintaining closely related monitoring
and verification variants of one MSR model creates a shared validation
point. Accepted observed traces conform to the monitorable variant within
the stated trust and abstraction boundaries, while documented
transformations connect that variant to the model used for proof. This
makes the remaining abstraction gap explicit and permits both variants to
be updated as the implementation evolves.
Third, our artifacts support model extraction for other Signal-based apps.

More broadly, monitorable models offer benefits beyond verification by enabling integration into the development
lifecycle for continuous validation. Developers can use these models as in-depth testing tools to
ensure ongoing compliance with protocol specifications and as a form of executable documentation.
Our work thus provides both the methodology and instrumentation for tech-savvy end users to establish greater
trust in their messaging applications, while offering vendors a systematic approach to document and maintain 
protocol correctness over time. 

All materials and models needed to reproduce and extend our results are available as described in~\cref{sec:open-science}.

%
%

\paragraph{Contributions}
We make the following contributions:
\begin{itemize}
	\item We are the first to demonstrate the feasibility of monitoring production-scale messaging applications with low measured overhead using SpecMon. We provide a methodology and reusable artifacts that enable quick and efficient monitoring of other Signal-based applications. 
  \item We develop the most detailed model of Signal, unifying and extending prior Tamarin models. We monitor it against the implementation and verify authentication and initial-root-key secrecy. Sealed Sender is monitored for fidelity but not separately verified.
  \item We develop the first formal model of WhatsApp Web's Signal-protocol variant, monitor it against the implementation, and verify authentication and initial-root-key secrecy. Both yield the known conversation-PCS counterexample.
\end{itemize}

\paragraph{Outline}
We introduce formal verification and runtime monitoring in \cref{sec:background}.
Our monitoring methodology is presented in \cref{sec:monitoring-methodology}, followed by case studies on Signal Desktop and WhatsApp Web in \cref{sec:signal-desktop,sec:whatsapp-web}.
We evaluate our approach through model validation, fault injection, and performance measurements in \cref{sec:evaluation}.
Formal verification results are discussed in \cref{sec:verification}, and related work in \cref{sec:related-work}.
Finally, we present limitations, lessons learned, future work, and conclusions in \cref{sec:limitations,sec:discussion,sec:conclusion}.

\section{Background}
\label{sec:background}

We first provide a high-level overview of the Signal protocol~\cite{signal-doc} used in both Signal Desktop and WhatsApp Web.
Then, we introduce SpecMon~\cite{specmon}, the runtime monitoring framework, and Tamarin~\cite{tamarin}, the verification tool used to formally verify the models.
Finally, we explain how symbolic rules become monitorable.

\subsection{The Signal Protocol}
\label{sec:background-signal}

The Signal protocol consists of two main subprotocols: an authenticated key agreement protocol, instantiated with the Extended Triple Diffie-Hellman (X3DH)~\cite{x3dh} or its post-quantum variant PQXDH~\cite{pqxdh}, and a continuous message encryption algorithm, instantiated with the Double Ratchet (DR)~\cite{dr}. The DR algorithm updates the key material used for sending and receiving messages after the initial key agreement. Typically, each pair of devices maintains one or more Signal connections (X3DH+DR). Each such connection is referred to as a session. Sesame~\cite{marlinspikeSesameAlgorithmSession2017} is the session management layer responsible for creating, deleting, and selecting these sessions.

At a high level, X3DH enables two parties to authenticate each other's identities while deriving a shared initial secret, called the initial root key. Its authentication and secrecy guarantees depend on which keys are compromised and when. PQXDH adds post-quantum forward secrecy, but its authentication still relies on the hardness of the discrete logarithm problem~\cite{pqxdh}. To enable asynchronous communication, each user uploads prekey bundles to the server, which include a hierarchy of keys: an identity key for the user, prekeys signed by the identity key, one-time prekeys, and, if supported, post-quantum prekeys signed by the identity key.

Using the initial secret as a seed, the DR algorithm derives message encryption keys while providing strong security guarantees: forward secrecy (FS)~\cite{DBLP:series/isc/BoydMS20}
ensures that past messages remain secure against future compromise, while post-compromise security (PCS)~\cite{DBLP:conf/csfw/Cohn-GordonCG16} allows future messages to become secure again after a current-state compromise and a subsequent healing period. The DR algorithm realizes these guarantees by combining
\begin{enumerate*}[label=(\alph*)]
  \item an asymmetric ratchet (PCS, FS) and
  \item a symmetric ratchet (FS).
\end{enumerate*}
In the asymmetric ratchet, parties exchange ephemeral Diffie-Hellman shares, and each new shared secret is merged into the root key. 
The symmetric ratchet uses a key derivation function (KDF) to derive message keys from the session secret. This way, every message is encrypted with a unique key, and compromise of later keys does not reveal information needed to compute earlier keys.

At the end-user level, Sesame manages the creation and use of sessions according to a specified policy, e.g., on decryption errors or desynchronization between the parties, which prevents some stronger PCS properties from holding~\cite{cremers2023a, DBLP:conf/sp/CremersMN25}.
Finally, the protocol can be extended to also provide sender anonymity by enveloping the Signal messages using the Sealed Sender algorithm~\cite{sealed-sender-doc}.

\subsection{SpecMon and Tamarin}
\label{sec:background-specmon-tamarin}

Tamarin and SpecMon use multiset-rewrite rules (MSRs) as their common specification language.
This common language lets one protocol model serve two purposes: Tamarin verifies security properties specified as trace properties, while SpecMon checks whether concrete implementation traces can be explained by the same rules.
Using the same specification language reduces divergence between the model used for verification and the model used for monitoring.

A rule is denoted as $\msrrewrite{l}{a}{r}$ and rewrites a multiset of \emph{facts}.
A fact consists of a fact symbol $\fact{F}$
and a list of terms.
The premise $l$, actions $a$, and conclusion $r$ are multisets of facts.
If the system state is described by a multiset of facts $S$, then the rule
is \emph{applicable} if $l \subseteq S$ and, in that case, its application
removes $l$ from $S$ and replaces it with $r$.
Such a transition is labeled with the actions $a$.
In Tamarin, terms are symbolic DY terms such as $\term{senc}(m,k)$, whose meaning is determined by user-defined equations.
Tamarin reasons about the symbolic traces generated by the MSRs in the presence of the DY attacker to verify the specified security properties.

SpecMon, by contrast, observes concrete bitstrings produced by an implementation.
It separates event extraction from checking: the event aggregator (EA) records relevant implementation events and forwards them to the monitor, which checks the event stream against the model.
This separation allows different event aggregation mechanisms depending on the deployment context.
We write such an event as $\langle f(x_1,\ldots,x_n), y\rangle$, where $f$ is the function name, the $x_i$ are concrete arguments, and $y$ is the concrete return value.

Since monitor events carry bitstrings rather than constructed terms, SpecMon rules cannot pattern-match on symbolic term structure in premises the way Tamarin rules can.
Instead, message structure is recovered through format strings, which pattern-match on bitstring layouts.
For cryptographic structure, the implementation must execute the corresponding function call. For example, instead of matching an encryption term in a premise, the monitor observes whether a decryption call succeeds.
The monitor maintains a set of configurations, each consisting of current facts.
If several rules can explain an event, SpecMon keeps the corresponding configurations until later events potentially resolve the nondeterminism.

MSRs enriched with monitoring annotations are called \emph{extended MSRs}.
SpecMon's soundness theorem~\cite[Theorem~7.1]{specmon} states that, for a \emph{likely} event stream (roughly, one in which freshly generated values are unique) accepted by a set of extended MSRs, there exists an abstraction from bitstrings to symbolic terms and a corresponding symbolic trace of the same model.
For properties stated over the modeled symbolic trace, this theorem provides the link from Tamarin proofs to accepted monitored executions within the modeled scope and assumptions. We expand on the threat model, trusted event extraction, and pre-trace initialization in \cref{sec:monitoring-methodology}.

\subsection{Monitorable Rules}
\label{sec:background-monitorable}

A Tamarin model is not immediately monitorable: implementations expose concrete events, randomness, return values, and bitstrings.
SpecMon therefore interprets function symbols in the symbolic model as function calls that the implementation may perform in the corresponding protocol state.
An observed event that no rule permits is reported as a violation.
SpecMon's rule decomposition~\cite[Def.~6.1]{specmon} splits a symbolic rule with nested function applications into a sequence of rules, each carrying at most one trigger for an individual function application.
Format strings add the information needed to parse and construct concrete messages, and trace rewriting optionally normalizes implementation-specific calling conventions and multi-call operations.

\paragraph{Triggers}
A trigger is a rule annotation.
Operationally, it is the observed program event that lets the monitor apply the rule, provided the rule's premise matches the current configuration.
Users can write triggers directly, while SpecMon provides them automatically for ordinary symbolic computations.
For example, the rule
\begin{lstlisting}
rule Hash:
  [Start(x)] --> [Next(h(x))]
\end{lstlisting}
is decomposed, schematically, into a rule with a trigger that matches an observed hash computation:
\begin{lstlisting}
rule Hash [x-trigger=[<h(x), y>]]:
  [Start(x)] --> [Next(h(x))]
\end{lstlisting}
When the event $\langle h(x), y\rangle$ is observed, SpecMon replaces the symbolic function application $h(x)$ with the concrete return value $y$ when applying the rule.
Similarly, facts $\fact{In}$ and $\fact{Out}$, which in Tamarin denote communication with the network attacker, are monitored through network I/O events, and $\fact{Fr}$, which asserts freshness, is monitored through randomness generation.

\paragraph{Format strings}
Network messages in a Tamarin model are symbolic terms, while implementations send and receive bitstrings.
Format strings specify how cryptographic messages are parsed and constructed.
%
%
For monitoring, we define a concrete bitstring layout with ciphertext length prefix $l$:
\[
  \begin{aligned}
    \mathtt{message}(RK, CT) ={}& \mathtt{cat}(\mathtt{string}(\texttt{'magic\_byte'}), \\
    & \mathtt{byte}(RK, 32), \mathtt{byte}(l, 1), \\
    & \mathtt{byte}(CT, l)),
  \end{aligned}
\]
For verification, the same symbolic message can be abstracted as a tuple giving the attacker access to each of its arguments:
\[
  \mathtt{message}(RK, CT) = \langle CT, RK \rangle.
\]
This lets SpecMon check bitstring-level messages while Tamarin reasons about symbolic terms.

\paragraph{Unified model}
SpecMon also supports Tamarin's preprocessor directives, including \texttt{\#ifdef}, \texttt{\#else}, and \texttt{\#endif}.
We use them to keep the verification-specific and monitoring-specific definitions in a single model file, for example to distinguish Tamarin's attacker and corruption rules from the concrete network and implementation events used during monitoring.

\paragraph{Trace rewriting}
Raw implementation events often differ from model events: function names may differ, functions may include extra implementation parameters, or one symbolic operation may be implemented through several library calls.
SpecMon supports trace rewriting rules that normalize these events before they are checked against the protocol model (see \cref{sec:trace-rewriting}).

\section{Monitoring Methodology}
\label{sec:monitoring-methodology}

\begin{figure}[t]
  \centering
  \includegraphics[width=\linewidth]{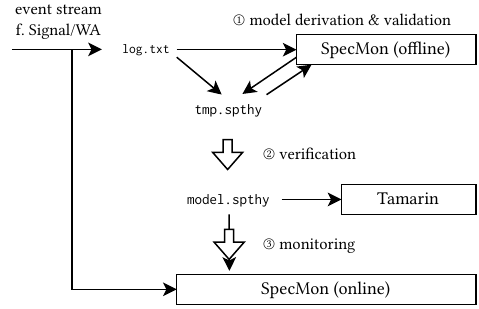}
  \caption{Methodology: data flow (solid) and outcomes (outlined).}
  \Description{Workflow for deriving, validating, verifying, and monitoring Signal and WhatsApp models online with SpecMon and Tamarin.}
  \label{fig:methodology}
\end{figure}

We describe how we extend SpecMon, instrument Signal Desktop and WhatsApp Web to extract relevant events, and adapt or derive models of the Signal protocol to fit the actual implementations. \Cref{fig:methodology} provides an overview of our methodology.

\paragraph{Threat model for monitoring}
The verification threat models in \cref{sec:verification} describe the
protocol attacker. Monitoring adds a deployment trust boundary.
We target \emph{honest-but-buggy} implementations and interpret SpecMon's
guarantees relative to trusted event extraction and pre-trace initialization,
both detailed below.
Concretely, we trust
\begin{enumerate*}[label=(\arabic*)]
  \item the instrumentation, which consists of annotated libraries for Signal Desktop and
    DevTools-based runtime wrappers for WhatsApp Web
    (\cref{sec:event-extraction}), together with the event aggregator,
  \item the browser's WebSocket and TLS stack, and
  \item the pre-trace initialization of the monitor's state
    (\cref{sec:state-initialization}).
\end{enumerate*}
Defenses against \emph{malicious} applications are out of scope and
discussed in \cref{sec:limitations}.

\paragraph{Scope of the assurance claim}
Monitoring ensures that accepted executions conform to the monitorable MSR model.
This guarantee is \emph{trace-relative} and depends on the trusted components above, the symbolic (Dolev-Yao) abstraction of cryptography, and the behaviors exercised during validation (\cref{sec:model-validation}).
Within this scope, monitoring catches observable model--implementation divergences, including incorrectly sequenced cryptographic operations, message-format mismatches, logical errors, invalid state-machine transitions, cryptographic misuse visible in events (e.g., a wrong IV), and implementation steps hidden by model abstractions.
However, it cannot detect behaviors outside the event stream, side channels, memory leaks, storage behavior, or errors on unexercised paths.

Two further aspects follow from the modeling language.
First, outputs are not enforced: the monitor checks that a sent message is \emph{permitted}, not that it is eventually sent.
Delaying or not sending a permitted message does not itself disclose additional data.
Second, Tamarin's input and freshness deduction applies in every state, so an implementation may sample additional randomness or receive additional messages without violating the model.
Unused randomness and inputs do not affect the modeled protocol behavior.
If these values are subsequently used in monitored operations, those operations must conform to the model.

\paragraph{Deployment scenarios}
We currently deploy monitoring for development, testing, and auditing: it checks an implementation's conformance to its formal model and can run as part of continuous integration, complementing protocol-level fuzzing.
Long-running production monitoring, for example in regulated domains or to produce audit evidence, is a natural extension. It requires automating the currently manual instrumentation, as discussed in \cref{sec:limitations}.

\subsection{Extending SpecMon}
\label{sec:extending-specmon}

Applying SpecMon to Signal Desktop and WhatsApp Web required extending the
monitor in two directions. First, we added support for protocol behavior that
appears in these implementations but was not covered by the previous monitor:
cryptographic computations whose return values are not reused later in the
protocol model, and recursive evaluation of nested format strings. These
extensions let the monitor describe the concrete encodings and helper
computations used by the applications without forcing artificial protocol state
into the model.

Second, we improved the monitor's execution engine to handle the size of these
case studies. In particular, we reduced repeated work in rule matching and state
updates, improved conflict-set computation to discard impossible states earlier,
and shared repeated representations of terms and facts in memory. These changes
do not alter SpecMon's monitoring semantics. They make the same modeling
approach practical for the larger rule sets, concurrent sessions, and event
streams encountered in our experiments. The extended SpecMon version
is included in the artifact.

%
%

\subsection{Event Extraction}
\label{sec:event-extraction}

The event aggregator (EA) is responsible for extracting relevant events from the application and feeding them to the monitor. This is done by instrumenting the application to log function calls to networking libraries and cryptographic components. We use two instrumentation strategies: annotated libraries when the relevant code is available, and dynamic browser instrumentation when it is not. The case-study sections describe the concrete instrumentation points for Signal Desktop and WhatsApp Web.

Network communication uses WebSockets protected by TLS. This means we trust the WebSocket implementation in the browser (for WhatsApp Web) or the Chromium networking stack bundled with Electron (for Signal Desktop). Ideally, rather than trusting the WebSocket layer, we would push the trust boundary down to the operating system's networking stack and monitor it there directly with SpecMon. This is possible in principle, but would require a holistic model covering both TLS and the protocol under consideration, which is currently infeasible due to the size and verification time of existing TLS models~\cite{cremersComprehensiveSymbolicAnalysis2017}.

Instrumenting the cryptographic library requires identifying its components and logging their function names, parameters, and return values.
SpecMon's guarantees depend on the EA emitting complete event streams, since it cannot distinguish a faulty implementation observed by a correct EA from a correct implementation observed by a faulty EA.
An omitted event may cause rejection if a later monitored rule depends on it, but this is not guaranteed.
For example, an uninstrumented send function may transmit data without the monitor observing it.
Omissions can therefore hide invalid behavior.
To allow a class of events to be omitted, one would have to show that adding those events back to an accepted stream preserves acceptance.

\subsection{State Initialization}
\label{sec:state-initialization}

One part of the protocol cannot be monitored: the \emph{setup}, typically the out-of-band configuration of common secrets.
SpecMon works around this issue by allowing monitoring to start with a \emph{pre-trace}, a prefix to the event stream. Typically, the user provides a script that reads the application's configuration and emits some specified event (e.g., $\langle \term{setup}('key'), \langle \rangle \rangle$) that the monitoring rules pick up to initialize facts.
Messaging apps retain sessions across restarts.
Their keys evolve locally (e.g., by ratcheting), and the updated session
state is persisted for the next run.
%

We use the pre-trace to recover this evolving state when the application starts using two scripts:
\begin{enumerate*}[label=(\arabic*)]
  \item one that extracts the relevant key material from the application's session table and
  \item one that emits events that initialize the monitor's state with these keys.
\end{enumerate*}

\subsection{Trace Rewriting}
\label{sec:trace-rewriting}

The event stream includes function calls as they appear in the library, often with implementation details outside the model.
Function names may differ from model symbols; for example, the library uses \texttt{aes\_encrypt} where the model uses \texttt{senc}.
Events may also include implementation-specific parameters (e.g., IVs and padding schemes) or split one model step into several calls, such as hash-state initialization, update, and finalization.

We use trace rewriting to adapt the event stream to the model.
This is done by specifying a set of rewriting rules that transform the event stream to match the model.
For example, we rewrite the function name \texttt{aes\_encrypt} to \texttt{senc} and combine multistep operations into a single step.
As shown in \cref{lst:trace-rewriting},
trace rewriting is built into SpecMon and uses the same MSR formalism as the protocol model:
the special action \texttt{PPEvent} passes program events to the next
layer, in our case, the Signal model.

\begin{lstlisting}[label={lst:trace-rewriting},
caption={Example: trace-rewriting rule that rewrites AES encryption calls to the model's symmetric encryption function. Here, \texttt{ct} denotes the ciphertext returned by \texttt{aes\_encrypt}.}]
rule ex [x-trigger=[<aes_encrypt(m,k,iv), ct>]]:
    [ ] --[PPEvent(<senc(m, k), ct>)]-> [ ]
\end{lstlisting}

\subsection{Model Construction}
\label{sec:model-construction}


When adapting a model for monitoring, we need to address abstractions of concrete message formats~\cite{modersheim2014, wallezComparseProvablySecure2023}, protocol features, and cryptographic operations.
This involves adding missing function calls and parameters observed in the traces, as well as adapting the model to account for differences in cryptographic operations.
For example, we extend \textcite{cremers2023a}'s Sesame model to include the derivation of a cipher key from the chain key, rather than encrypting directly with the chain key.
We follow a systematic approach to adapt the model based on the rewritten traces from the EA: we analyze the cause of a rejected event, verify its validity according to the specification, and adapt the model accordingly.

Monitoring also revealed details that our initial adaptation of the Sesame model had omitted.
The original model abstracts session setup by initializing a chain key directly from a fresh root key.
In Signal Desktop, the initiator first performs a Diffie-Hellman ratchet step using its fresh ratchet key and the recipient's signed prekey.
This step updates the root key and derives the sending chain key before the first message is encrypted.
SpecMon rejected the corresponding trace because our initial adaptation lacked this step.
We therefore extended the model to include it.

\paragraph{Missing format strings}
    Input and output events that cannot be matched may indicate missing or incorrect format strings in the model.
    This is to be expected, as verification models often abstract away message formats as a simple nested pair encoding.
    We consult the Signal specification~\cite{signal-doc} and implementation~\cite{signal-d} for details, noting that Signal uses Protocol Buffers~\cite{storageproto} for its wire format.

\paragraph{Incorrect transition}
    SpecMon outputs the current facts when it rejects an event.
    If that state is not the one expected at this point in the trace, we identify the rule that led to the observed state fact and correct it.
    If there are multiple current states
    or if the prior state was also incorrect, we truncate the trace to find the first rejected event and inspect the state at that point to analyze what led us to the current state.

\paragraph{No valid transitions} In this case, we check:
    \begin{itemize}[beginpenalty=10000]
      \item If there is a \emph{missing transition} to an existing state, we add a new rule for this transition.
      \item If the model expects a \emph{different order of operations} than the implementation, we adapt the model accordingly.
      \item If the model is \emph{incomplete}, we either extend the model by adding missing functionality
        or introduce abstractions. 
    \end{itemize}

\section{Signal Desktop}
\label{sec:signal-desktop}

We target the desktop version of Signal, written in TypeScript.
We monitor the PQXDH key-exchange protocol, the DR protocol, and the Sealed Sender mechanism in Signal Desktop~\cite{signal-d}.
Since the implementation is open-source, we adopt the annotated-library approach from SpecMon~\cite{specmon}.
For both network and cryptographic components, we annotate functions in \libsignal{}~\cite{libsignal} used by Signal Desktop~\cite{signal-d},
and use our modified versions to run the app.

Furthermore, we monitor the session initialization in Signal Desktop by merging the Sesame model~\cite{cremers2023a} with the X3DH model~\cite{albert}. The model is then extended to fit the actual implementation.

\subsection{Event Aggregation}
\label{sec:signal-ea}

Network events come from the WebSocket handlers \texttt{onmessage} (incoming) and \texttt{send\_request} (outgoing).
For cryptography, Signal clients and servers use the platform-agnostic \libsignal{}~\cite{libsignal} library.
It is implemented in Rust and exposed to JavaScript.
We annotate the functions that appear in the base models for DR~\cite{cremers2023a} and X3DH~\cite{albert}, as listed in
\iffullversion\cref{tab:signal-instrumented-functions}\else\cref{tab:instrumented-functions}\fi.

\subsection{Pre-traces and Database Decryption}
\label{sec:signal-pretraces}

Signal Desktop stores session keys in an encrypted database on the user's device.
On Linux, it is typically located in the user's application configuration directory (e.g., \verb|~/.config/Signal| for production or \verb|~/.config/Signal-development| for staging).

To obtain the database decryption key, we read the Signal profile configuration and recover the SQLCipher key from the stored \texttt{key} or \texttt{encryptedKey} entry.
On Linux, this requires decrypting the stored key using the secret retrieved from the system secret service. We then use the resulting SQLCipher key to 
decrypt the database, extract the current session keys, and provide them to the monitor as pre-traces.
For each session, we extract the root key; the sending ratchet key pair and chain key;
the receiving public ratchet keys and chain keys; the base key (the public key of the
ephemeral secret used in the handshake); and the session identifier (a tuple of the
sender's and recipient's identifiers).
These pre-traces act as triggers for rules that have no premises and whose
conclusions introduce the facts required by subsequent rules. This setup allows us
to monitor pre-existing sessions of Signal Desktop without having to monitor session
initialization every time (see \cref{sec:state-initialization}).

\subsection{Trace Model Construction}
\label{sec:signal-model-construction}

We start with the DR model by \textcite{cremers2023a} for monitoring the DR protocol in Signal Desktop, and with the X3DH model~\cite{albert} for the handshake.
We then merge both models and extend them to fit the implementation traces and include missing details. These extensions cover post-quantum ML-KEM keys used during session initialization, track ratchet keys, and model the derivation of message and cipher keys from chain keys for encryption and decryption.
We explicitly model the initial DH ratchet step before the first message and the successive root-key updates that derive a party's receiving and sending chains.

Additionally, Signal Desktop uses a sender-identity-hiding mechanism, called Sealed Sender.
This mechanism is used on top of the DR and PQXDH protocols to hide the sender's identity from the server.
Since neither initial model covers Sealed Sender and no prior model is available,
we extend the model with rules for this mechanism.
This makes our final model the most detailed model of Signal to date.
Our model includes the following rules.
\begin{itemize}
  \item Initiator: starts a session with a recipient.
  \item Responder: receives the initial message and responds.
  \item Prekey: the recipient publishes a prekey bundle to the server.
  \item Sender symmetric ratchet: advances the sending chain key and encrypts a message.
  \item Receiver symmetric ratchet: advances the receiving chain key and decrypts a message.
  \item Asymmetric (Diffie-Hellman) ratchet: updates the root key and derives new sending and receiving chain keys when receiving a message with a new ratchet key.
  \item Sealed send: the sender encrypts their public identity key using the recipient's public key and an ephemeral key, and encrypts the message using the recipient's public identity key together with the sender's private identity key.
  \item Sealed receive: the recipient decrypts the sender's identity using their private identity key and the ephemeral key, and the message using the sender's public identity key and their own private identity key.
\end{itemize}

\subsection{Format Strings}
\label{sec:signal-format-strings}

We extend the model with format strings to cover the actual structure of messages as observed in the traces. We derive the message structure from the Protocol Buffers definitions used in Signal Desktop~\cite{storageproto}, following the approach described in \cref{sec:background-monitorable}.

\section{WhatsApp Web}
\label{sec:whatsapp-web}

WhatsApp Web~\cite{wa} is a closed-source application, which makes monitoring more challenging than monitoring Signal Desktop~\cite{signal-d}. However, although its JavaScript source code is minified, it remains partially readable, and many function names align closely with those from the white paper~\cite{whitepaper}.
This correspondence significantly simplifies the task of identifying functions and operations relevant to the Signal protocol.

\subsection{Event Aggregation}
\label{sec:whatsapp-event-aggregation}

WhatsApp Web communicates over WebSockets. We use Chrome DevTools to identify and override WebSocket-related functions. To distinguish calls relevant to the Signal protocol from unrelated traffic, we inspect the stack trace at the time of the WebSocket invocation. The stack trace reveals the calling context, allowing us to selectively forward relevant events to SpecMon.

Additionally, since WhatsApp Web runs within the browser sandbox, it can only communicate over browser-exposed APIs. This security model simplifies monitoring by ensuring that the application cannot bypass our instrumentation through lower-level channels (e.g., using system calls, as would be possible in a native application). We assume that the browser and TLS are trusted and extract network I/O
from the WebSocket components (\cref{sec:event-extraction}).

WhatsApp Web uses a modified version of \libsignal{}~\cite{libsignal} for its cryptographic operations~\cite{WAlibsignal}. We identify cryptographic components by matching function names observed at runtime with those described in the white paper~\cite{whitepaper}. Using Chrome DevTools, we intercept these functions and step into them via the debugger to analyze their logic in detail. We then override the relevant cryptographic functions, as shown in
\iffullversion\cref{tab:whatsapp-instrumented}\else\cref{tab:instrumented-functions}\fi,
to log event traces to standard output for SpecMon to consume.


\subsection{Pre-traces and Database Access}
\label{sec:whatsapp-pretraces}

WhatsApp Web stores session keys in the IndexedDB database \texttt{signal-storage}.
For each session, the \texttt{session-store} object store holds the base key; the root key;
the five most recent receiver chains (including the receiver's public ratchet keys
and chain keys); and the sender's chain (including the sender's most recent ratchet
key pair and chain key).

Other tables, like \texttt{prekey-store} and \texttt{identity-store}, contain the user's one-time keys and public identity key.
However, the private identity key is not saved in the database---it is stored as a non-extractable \texttt{CryptoKey} object in the browser's memory and is not directly accessible.

{\setlength{\emergencystretch}{2em}%
Using Chrome DevTools, we query the database to extract the available keys. For the private identity key, we instead access the code module that loads it during runtime through the internal registration API
\texttt{WAWebCryptoLibrary.}\allowbreak\texttt{DbCallbacksApi.}\allowbreak\texttt{getRegistrationInfo}.
We then pass all these keys to the monitor as pre-traces (\cref{sec:state-initialization}).
\par}

\subsection{Model Derivation}
\label{sec:whatsapp-model-derivation}

For WhatsApp Web, no prior formal model is available. We derive a model of its implementation of the Signal protocol directly from the recorded traces.
The model covers both the Double Ratchet protocol and the X3DH key exchange protocol.
Our model includes the following rules.
\begin{itemize}
  \item X3DH initiator: starts a session with a recipient.
  \item X3DH responder: receives the initial message and responds.
  \item X3DH prekey: publishes the recipient's bundle to the server.
  \item Sender symmetric ratchet: advances the sending chain key and encrypts a message.
  \item Receiver symmetric ratchet: advances the receiving chain key and decrypts a message.
  \item Asymmetric (Diffie-Hellman) ratchet: updates the root key and derives new sending and receiving chain keys when receiving a message with a new ratchet key.
\end{itemize}

\subsection{Format Strings and BLOB Decoding}
\label{sec:whatsapp-format-strings}

When the user sends a message (by clicking Send in the UI), WhatsApp Web encrypts it for both the user's primary device (phone) and all recipient devices. Each encrypted payload is serialized as a separate \texttt{SignalMessage}, which contains the ciphertext, public ratchet key, and counters using Protocol Buffers~\cite{protobuf}. These individual \texttt{SignalMessage} objects are then bundled together into a single \texttt{WebSocketMessage} for transmission.

Because a single WebSocket BLOB can contain an arbitrary number $n$ of \texttt{SignalMessage} objects, the monitor would require a format string for every value of $n$. To manage this complexity, we delegate the decoding of the BLOBs and \texttt{SignalMessage} objects to the event aggregator (EA). Using Chrome DevTools, the EA extracts and decodes each \texttt{SignalMessage} from the WebSocket bundle and emits it individually to the monitor. Consequently, the monitor receives each \texttt{SignalMessage} as a separate event, allowing it to process the messages one by one.

\subsection{Findings: Differences from Signal Desktop}
\label{sec:whatsapp-finding-read-receipts}

Monitoring surfaced two differences between WhatsApp Web and Signal Desktop.
The first is an engineering difference whose theoretical bearing on security
we discuss below, and the second is a known feature-adoption difference.

\paragraph{Observation 1: read receipts outside the Double Ratchet}
We observed that WhatsApp Web transmits read-receipt messages outside the Double Ratchet encryption layer, unlike text messages and reaction messages. In contrast, Signal Desktop integrates such messages into the DR protocol. This distinction is evident during runtime monitoring. Using SpecMon, we observe the DR rules being triggered and applied in Signal Desktop even for read-receipt messages, whereas in WhatsApp Web they are triggered and applied only for actual message exchanges. Furthermore, inspecting the \texttt{signal-storage} database in WhatsApp Web during read-receipt transmission shows no updates to any of the keys, confirming that these events are not encrypted as part of the DR protocol.
Instead, WhatsApp Web handles read receipts on the server side.
The client sends a server-visible \emph{presence message} whenever the user opens a chat. Depending on the user's settings, the server decides whether to send a read receipt to the other party.

As a result, we observe that WhatsApp Web performs DH ratchets less frequently than Signal Desktop.
In Signal Desktop, read receipts participate in the encrypted message exchange and can trigger DH ratchets. In WhatsApp Web, receiving a read receipt does not trigger a DH ratchet.

For a single session, post-compromise recovery depends on the asymmetric
ratchet mixing fresh DH material into the root key. Signal may therefore
heal sooner than WhatsApp for comparable user interactions. Forward secrecy
for already-sent messages is provided by the symmetric ratchet in the modeled
protocol and is not weakened by this difference in DH-ratchet frequency.
In practice, we do not conclude a weakening of WhatsApp's security from this
observation alone. Whether a compromised session actually heals also depends
on how long old key material persists in memory and storage, which our
methodology does not observe. We show no concrete attack based on this
difference. An analysis of finer-grained healing properties remains future work.

\paragraph{Observation 2: feature-adoption differences}
The WhatsApp Web version studied here does not incorporate a post-quantum mechanism (such as
Signal's PQXDH) during session initialization and has no mechanism to hide
the sender's identity (such as Signal's Sealed Sender), unlike Signal
Desktop. These are known feature-adoption differences between the
applications. Our contribution is
to surface them at the model level and document their behavioral consequences
through monitoring.

%

\section{Evaluation}
\label{sec:evaluation}

We performed an extensive evaluation of our monitoring approach on both Signal Desktop~\cite{signal-d} and WhatsApp Web~\cite{wa}.


\subsection{Model Validation}
\label{sec:model-validation}

To validate the functionality of our monitoring models for Signal Desktop and WhatsApp Web,
we monitor a party \emph{Monique}, running the respective software,
in communication with an unmonitored partner \emph{Parker}.
Our goal is to cover a variety of bilateral communication patterns,
including network and state anomalies. We do not attempt to cover
group messages (not covered by our models) 
or linked
devices for Signal (from the protocol perspective, linked devices are essentially
independent communication partners).

To this end, we trigger the following UI or external actions,
which capture the main behaviors of the
Signal protocol: 
\begin{enumerate*}[label=(\arabic*)]
  \item \label{it:new} starting a new conversation, triggering session \emph{initialization};
  \item \label{it:send} sending $m$ consecutive messages, triggering \emph{symmetric} ratcheting;
  \item \label{it:receive} receiving $m$ consecutive messages, triggering \emph{symmetric} ratcheting;
  \item \label{it:net-swap} swapping two messages in the network stack, so they appear out of order;
  \item \label{it:net-skip} skipping a message in the network stack; and
  \item \label{it:session-loss} sending or receiving a message to or from a device that has lost its session, triggering a \emph{retry request}. This can occur when the communication partner switches phones or otherwise loses state.
\end{enumerate*}

For WhatsApp Web,
\ref{it:net-swap} and \ref{it:net-skip} intercept messages after Signal-protocol
encryption and before Noise transport encryption. Reordering the encrypted
WebSocket frames instead would violate Noise's sequence-number checks.
Both our models and instrumentation support out-of-order messages, and the
artifact includes the corresponding fuzzer actions.
See \cref{sec:eval-instrumentation} for details.

The artifact provides randomized fuzzers that exercise the enabled actions.
The Signal Desktop fuzzer selects uniformly among actions whose prerequisites hold.
The WhatsApp Web fuzzer first selects uniformly among the available action groups---sending,
receiving, session initialization, retry, and out-of-order delivery when enabled---and then
uniformly within the selected group. Actions that replay held messages are
eligible only when such messages are available. Consecutive-send/receive batch
sizes use a configurable rounded lognormal distribution with a minimum of one.
The action set, distribution parameters, and run length are configurable.
The idea is that the actions cover all the features in the scope of
the monitor so that, from their random arrangement, corner cases may
emerge that we may not anticipate with handwritten tests.
For instance, while we knew a~priori that
switching from sending to receiving (or vice versa) would
trigger the \emph{Diffie-Hellman} ratcheting step, we found that 
the
combination of \ref{it:session-loss}
followed by \ref{it:send} triggers fail-safe behavior.
When Parker
receives Monique's message for the deleted session,
they cannot decrypt it and return a decryption error. Monique then starts
a new session and sends a new message.

We validate the models on three mixed workloads.
We count selected send/receive instrumentation observations, not necessarily
distinct chat messages.
The Signal Desktop trace contains 995 send/receive observations and 35,040 raw
instrumentation observations (\qty{25.02}{\mega\byte}). The ordinary WhatsApp Web trace
contains 418 send/receive observations and 11,918 raw observations (\qty{6.51}{\mega\byte}); its out-of-order workload
contains 499 send/receive observations and 15,960 raw observations (\qty{8.47}{\mega\byte}).
Rewriting and monitoring these traces yields 17,393, 7,232, and 9,019 monitor
events, respectively, including initialization events. All replays succeed.

We also replay consecutive-send workloads containing 7,007 send observations for
Signal Desktop and 3,009 for WhatsApp Web, with 168,271 and 73,246 raw
observations, respectively. Both replays succeed.
Additional ratcheting traces exercise repeated alternation between sending and
receiving, and these replays also succeed.

\subsection{Fault Injection Experiments}
\label{sec:fault-injection}

To evaluate the effectiveness of monitoring in detecting security-critical errors,
we inject faults through implementation modifications and a trace-level omission.
SpecMon detects the tested protocol-level faults; only payload-encoded leakage is outside the model's reach.

\subsubsection*{Critical secret leakage} We leak a critical secret in three ways.
First, we send it in an extra message.
Second, we place it in a chat message's content.
Third, we add it as a new field in a protocol message.

In the first case, SpecMon detects the leak because of an unexpected event in the trace, and in the third case, because of an unmatched format string.
In the second case, SpecMon cannot distinguish the secret from a legitimate message, since the secret is encoded as application payload.
See~\cref{sec:limitations} for details.

\subsubsection*{Incorrect use of a cryptographic library}
First, during the Diffie-Hellman ratchet, we reuse a DH private key in
Signal Desktop. For WhatsApp Web, we omit the key-generation event from the
trace. SpecMon rejects the subsequent Diffie-Hellman computation because the
required fresh-key generation was not observed.
Second, we skip a signature check in the function
\texttt{process\_prekey\_bundle}, so all prekey bundles from the server are
accepted (similar to goto-fail, CVE-2014-1266~\cite{gotofail}).
SpecMon detects the missing signature verification in both applications.
These experiments do not establish the random-number generator's security, as predictability of fresh-looking random values is outside the symbolic model.

\subsection{Performance Overhead}
\label{sec:performance-overhead}

\begin{figure*}[tb]
    \newcommand{\myheight}{50mm}
    \centering
    \begin{subfigure}[t]{0.3\textwidth}
        \resizebox{\myheight}{!}{
\begin{tikzpicture}

\definecolor{color0}{rgb}{0.0729412,0.28,0.423529}
\definecolor{color1}{rgb}{0.103529,0.376471,0.103529}

\begin{axis}[
height=6.3cm,
legend cell align={left},
legend style={fill opacity=0.9, draw opacity=1, text opacity=1, at={(0.03,0.97)}, anchor=north west, draw=white!80!black},
minor xtick={},
minor ytick={},
tick align=outside,
tick pos=left,
width=8.0cm,
x grid style={white!69.0196!black},
xlabel={Send/receive observations (Signal / WhatsApp)},
xmajorgrids,
xmin=40, xmax=950,
xtick style={color=black},
xtick={90,180,270,360,450,540,630,720,810,900},
xticklabels={\shortstack{90\\40},\shortstack{180\\80},\shortstack{270\\120},\shortstack{360\\160},\shortstack{450\\200},\shortstack{540\\240},\shortstack{630\\280},\shortstack{720\\320},\shortstack{810\\360},\shortstack{900\\400}},
y grid style={white!69.0196!black},
ylabel={Avg. time/event (\unit{\milli\second})},
ymajorgrids,
ymin=0, ymax=0.75,
ytick style={color=black},
ytick={0,0.25,0.5,0.75}
]
\addplot [line width=1.8pt, color0, mark=*, mark size=1.9pt, mark options={solid, fill=white}]
table {%
90 0.315157
180 0.336150
270 0.297840
360 0.310904
450 0.281109
540 0.287972
630 0.291014
720 0.284107
810 0.277486
900 0.295531
};

\addlegendentry{Signal}

\addplot [line width=1.8pt, color1, mark=square*, mark size=1.8pt, mark options={solid, fill=white}]
table {%
90 0.082120
180 0.099550
270 0.109081
360 0.123200
450 0.148895
540 0.162111
630 0.170469
720 0.197519
810 0.229084
900 0.256930
};

\addlegendentry{WhatsApp}
\addplot [line width=1.5pt, orange!80!black, dashed, mark=triangle*]
table {%
90 0.098848
180 0.135215
270 0.193410
360 0.206690
450 0.231542
540 0.204600
630 0.238472
720 0.267988
810 0.258925
900 0.271264
};
\addlegendentry{WhatsApp OOO}
\end{axis}

\end{tikzpicture}}
        \caption{Avg. processing time per event.}
        \label{fig:performance-processing-time}
    \end{subfigure}
    \begin{subfigure}[t]{0.3\textwidth}
        \resizebox{\myheight}{!}{
\begin{tikzpicture}

\definecolor{color0}{rgb}{0.0729412,0.28,0.423529}
\definecolor{color1}{rgb}{0.103529,0.376471,0.103529}

\begin{axis}[
name=memtop,
height=3.915cm,
legend cell align={left},
legend style={fill opacity=0.9, draw opacity=1, text opacity=1, at={(0.03,0.97)}, anchor=north west, draw=white!80!black},
minor xtick={},
minor ytick={},
tick align=outside,
tick pos=left,
width=8.0cm,
x grid style={white!69.0196!black},
xmin=40, xmax=950,
xtick style={color=black},
xtick={90,180,270,360,450,540,630,720,810,900},
xticklabels={90,180,270,360,450,540,630,720,810,900},
y grid style={white!69.0196!black},
ylabel={Peak RSS (\unit{\mebi\byte})},
ymajorgrids,
ymin=0, ymax=96,
ytick style={color=black},
ytick distance=10,
tick label style={font=\small},
xticklabel style={font=\small}
]
\addplot [line width=1.8pt, color0, mark=*, mark size=1.9pt, mark options={solid, fill=white}]
table {%
90 31.285156
180 31.683594
270 53.313802
360 39.002604
450 43.255208
540 48.410156
630 49.544271
720 50.061198
810 52.834635
900 53.537760
};

\addlegendentry{Signal}
\end{axis}

\begin{axis}[
name=membottom,
at={(memtop.south west)},
anchor=north west,
yshift=-0.70cm,
width=8.0cm,
height=3.915cm,
legend cell align={left},
legend style={fill opacity=0.9, draw opacity=1, text opacity=1, at={(0.97,0.03)}, anchor=south east, draw=white!80!black},
minor xtick={},
minor ytick={},
tick align=outside,
tick pos=left,
xmin=40, xmax=950,
xtick style={color=black},
xtick={90,180,270,360,450,540,630,720,810,900},
xticklabels={40,80,120,160,200,240,280,320,360,400},
x grid style={white!69.0196!black},
y grid style={white!69.0196!black},
ylabel={Peak RSS (\unit{\mebi\byte})},
ymajorgrids,
ymin=0, ymax=50,
ytick style={color=black},
ytick distance=10,
xlabel={Send/receive observations},
tick label style={font=\small},
xticklabel style={font=\small},
xlabel style={font=\small}
]
\addplot [line width=1.8pt, color1, mark=square*, mark size=1.8pt, mark options={solid, fill=white}]
table {%
90 21.996094
180 24.725260
270 27.216146
360 27.052083
450 29.302083
540 29.052083
630 32.257812
720 35.964844
810 37.440104
900 35.240885
};

\addlegendentry{WhatsApp}
\addplot [line width=1.5pt, orange!80!black, dashed, mark=triangle*]
table {%
90 26.223958
180 29.013021
270 29.059896
360 33.406250
450 30.127604
540 35.119792
630 33.074219
720 37.023438
810 39.384115
900 43.304688
};
\addlegendentry{WhatsApp OOO}
\end{axis}

\end{tikzpicture}}
        \caption{SpecMon process memory usage.}
        \label{fig:performance-memory}
    \end{subfigure}
    \begin{subfigure}[t]{0.3\textwidth}
        \resizebox{\myheight}{!}{
\begin{tikzpicture}

\definecolor{color0}{rgb}{0.12156862745098,0.466666666666667,0.705882352941177}
\definecolor{color1}{rgb}{1,0.498039215686275,0.0549019607843137}

\begin{axis}[
legend cell align={left},
legend style={fill opacity=0.8, draw opacity=1, text opacity=1, draw=white!80!black},
tick align=outside,
tick pos=left,
x grid style={white!69.0196078431373!black},
xlabel={Messages},
xmajorgrids,
xmin=-1.45, xmax=52.45,
xtick style={color=black},
xtick={0,10,20,30,40,50},
y grid style={white!69.0196078431373!black},
ylabel={Average latency (\unit{\milli\second})},
ymajorgrids,
ymin=2, ymax=7,
ytick style={color=black},
ytick distance=1
]
\path [fill=color0, fill opacity=0.25]
(axis cs:1,5.83245553203)
--(axis cs:1,4.56754446797)
--(axis cs:5,3.13017246508)
--(axis cs:10,3.12652907804)
--(axis cs:15,2.94949956123)
--(axis cs:20,2.86551663562)
--(axis cs:25,2.91453768472)
--(axis cs:30,2.85743213465)
--(axis cs:35,2.90431899892)
--(axis cs:40,2.85391258918)
--(axis cs:45,2.94347743253)
--(axis cs:50,2.87404019342)
--(axis cs:50,3.24995980658)
--(axis cs:45,3.25652256747)
--(axis cs:40,3.11108741082)
--(axis cs:35,3.34710957251)
--(axis cs:30,3.12923453201)
--(axis cs:25,3.24546231528)
--(axis cs:20,3.21448336438)
--(axis cs:15,3.22383377211)
--(axis cs:10,3.85347092196)
--(axis cs:5,3.70982753492)
--(axis cs:1,5.83245553203)
--cycle;

\path [fill=color1, fill opacity=0.25]
(axis cs:1,5.41622776602)
--(axis cs:1,4.78377223398)
--(axis cs:5,2.80420090321)
--(axis cs:10,2.7032143009)
--(axis cs:15,2.59320990375)
--(axis cs:20,2.60930058036)
--(axis cs:25,2.65501322835)
--(axis cs:30,2.53475714088)
--(axis cs:35,2.71192905987)
--(axis cs:40,2.76366647856)
--(axis cs:45,2.65700529489)
--(axis cs:50,2.65767145911)
--(axis cs:50,2.86632854089)
--(axis cs:45,2.91632803844)
--(axis cs:40,2.91133352144)
--(axis cs:35,2.94521379728)
--(axis cs:30,3.03190952579)
--(axis cs:25,2.86498677165)
--(axis cs:20,3.07069941964)
--(axis cs:15,3.15345676292)
--(axis cs:10,3.6367856991)
--(axis cs:5,3.63579909679)
--(axis cs:1,5.41622776602)
--cycle;

\addplot [semithick, color0]
table {%
1 5.2
5 3.42
10 3.49
15 3.08666666667
20 3.04
25 3.08
30 2.99333333333
35 3.12571428571
40 2.9825
45 3.1
50 3.062
};
\addlegendentry{with instrumentation}
\addplot [semithick, color1]
table {%
1 5.1
5 3.22
10 3.17
15 2.87333333333
20 2.84
25 2.76
30 2.78333333333
35 2.82857142857
40 2.8375
45 2.78666666667
50 2.762
};
\addlegendentry{without instrumentation}
\end{axis}

\end{tikzpicture}}
        \caption{Avg. latency per message in Signal Desktop.}
        \label{fig:performance-latency}
    \end{subfigure}
    \caption{SpecMon performance for Signal Desktop and WhatsApp Web. Bands in (\subref{fig:performance-latency}) show mean $\pm$ one sample standard deviation across run means.}
    \Description{Three plots showing average processing time per event, memory usage, and Signal Desktop latency while monitoring Signal Desktop and WhatsApp Web. The shaded latency bands show one sample standard deviation across run means above and below the mean latency.}
    \label{fig:performance}
\end{figure*}
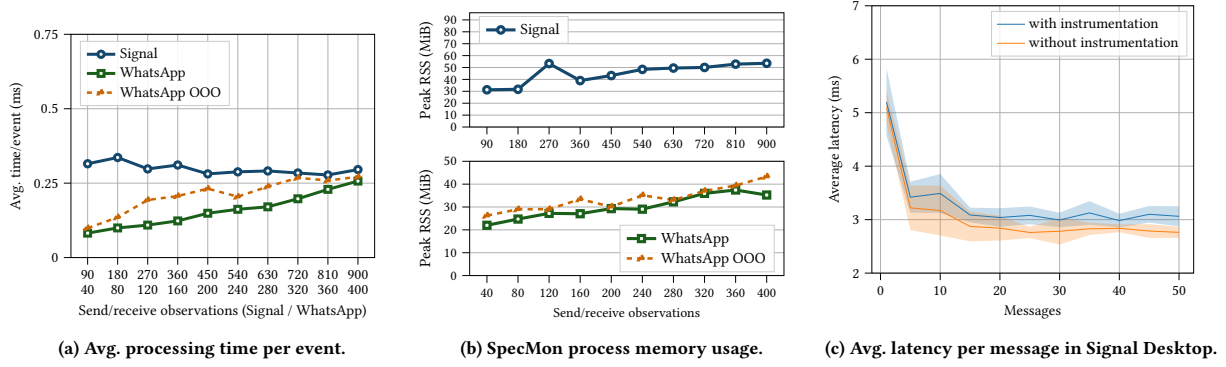

We measure monitoring performance by replaying three workloads:
Signal Desktop, WhatsApp Web, and WhatsApp Web with out-of-order delivery.
We use a native Linux ARM64 virtual machine
with 16 virtual CPUs and \qty{16}{\gibi\byte} of RAM, Go 1.26.8, and three sequential
repetitions per prefix after one warmup at the smallest prefix size.
Prefixes contain 90, 180, \ldots, 900 send/receive observations for Signal and
40, 80, \ldots, 400 for each WhatsApp workload.
These observations need not represent distinct messages.

\Cref{fig:performance-processing-time,fig:performance-memory} show the
mean main-monitor processing time per event and mean peak process RSS.
The processing-time statistic excludes the preceding rewrite stage; peak RSS
covers the entire SpecMon process, including rewriting. The ranges are
Signal: \qtyrange{0.277}{0.336}{\milli\second} per event and \qtyrange{31.29}{53.54}{\mebi\byte} peak RSS;
WhatsApp: \qtyrange{0.082}{0.257}{\milli\second} per event and \qtyrange{22.00}{37.44}{\mebi\byte} peak RSS;
WhatsApp OOO: \qtyrange{0.099}{0.271}{\milli\second} per event and \qtyrange{26.22}{43.30}{\mebi\byte} peak RSS.
Across individual repetitions, peak RSS reaches \qty{83.56}{\mebi\byte} for Signal,
\qty{43.99}{\mebi\byte} for WhatsApp, and \qty{46.09}{\mebi\byte} for WhatsApp with out-of-order delivery.

The Signal prefix with 900 send/receive observations contains 31,692 raw observations and 15,724 monitor
events, including its initialization pretrace. The WhatsApp prefix with 400 send/receive observations
contains 11,446 raw observations and 6,938 monitor events; the corresponding
WhatsApp out-of-order prefix contains 12,915 raw observations and 7,195 monitor
events. Thus the ordinary prefixes contain 35.21 and 28.62 raw observations per
send/receive observation, respectively.

In a separate experiment, we evaluate the latency of Signal Desktop.
The overall latency consists of SpecMon's processing time and
the computation overhead of the instrumentation (e.g.,
dereferencing pointers, copying buffers, and building strings in the
event streams).
We only measure end-to-end latency for Signal Desktop, as we instrument WhatsApp Web
dynamically at runtime (`monkey patching'), which is not a realistic
deployment scenario. The expectation that source-level WhatsApp
instrumentation would have comparable latency is therefore an extrapolation
from the comparable SpecMon processing times
(\cref{fig:performance-processing-time}).
Monkey patching identifies functions by name in the minified bundle,
which makes it a cheap alternative to reverse-engineering a closed-source
client from scratch; in exchange, each upstream rename or bundle
reshuffle requires relocating the affected functions.
For Signal Desktop, we define latency as the time between the event of inputting a message
(collected from the function \texttt{sendMessageToServiceId} in~\cite{signal-d}) and the
event of the encrypted payload being handed off to the WebSocket layer for transmission
(collected from \texttt{sendMessages} in~\cite{signal-d}), i.e., the encryption overhead of the send path.
For each $m$, we measure the latency on the unaltered code (the
baseline) and on the
instrumented code interacting with SpecMon.
In \cref{fig:performance-latency}, we report both latency measurements over $m\in\{1,5,10,15,\dots,50\}$.
The curves flatten by $m=50$, so we omit larger values.
We find that the latency differences fall within the fluctuation of the
average latency. This variation decreases as the number of messages increases. For $m=50$, the average instrumentation overhead is \qty{0.30}{\milli\second} over the uninstrumented average of \qty{2.76}{\milli\second}.
This performance can still be improved upon, as
the prototype is not particularly optimized. For example, the instrumentation sends JSON objects to SpecMon through standard output.
Calling the SpecMon library directly would avoid several copy operations.

\section{Verification}
\label{sec:verification}

In this section, we summarize our verification results for the core key-agreement and session components of the \whatsapp and \signal models.
The monitorable \signal model additionally covers Sealed Sender for
implementation fidelity, but we do not verify a sender-privacy property for
it. This would require an observational-equivalence analysis outside our current
Tamarin workflow.

\subsection{Models}
\label{sec:verification-models}
Our models are structured around three components: the setup phase, the key agreement, and the message exchange.

\paragraph{Setup} We model an unbounded number of users that can have an unbounded number of sessions and send or receive an unbounded number of messages during their communication. Each user is equipped
with an unbounded number of prekey bundles, and depending on whether they execute X3DH or PQXDH, we model different initialization steps. In \whatsapp, we model a long-term identity key pair
\z{(\mathit{ltk}_A, \ltka)}, signed prekeys
\z{(\sk_A, \SK_A)}, and one-time prekeys \z{(\opk_A, \OPK_A)}.
The numbers of signed and one-time prekeys are unbounded.
For \signal, we also model an unbounded number of one-time post-quantum keys \z{(\kybersk_A, \kyberPK_A)}.

\paragraph{Key Agreement} As discussed, \signal employs the PQXDH protocol and \whatsapp the classical X3DH protocol. 
Typically, modelers would condense the key agreement into two rules for the initiator and responder, since this helps during verification. However, to allow for monitoring, our model represents the
step-by-step atomic construction of the state in the implementation, where, for example, the initial key can be computed only after verification of the prekey signatures.
We capture the key agreement for \whatsapp
in six initiator rules and two responder rules, and for \signal in five initiator rules and two responder rules.

\paragraph{Message Exchange} The Double Ratchet protocol is defined by its two components: the symmetric and asymmetric ratchets. In the symmetric ratchet, captured in three rules, each party can send and receive an unbounded number of messages by advancing the sending and receiving chain keys. This is expressive enough to capture out-of-order delivery of messages. The asymmetric ratchet models the updated state upon receiving a message with a new Diffie-Hellman share. In contrast to previous models, ours
captures the implementation's fast-forwarded state, where an incoming message immediately
also advances the sending chain, i.e., if $B$ receives a message, they consecutively perform two asymmetric-ratchet steps: one to update the receiving chain and decrypt the message, and the other to prepare a new sending chain for the future. This is captured in two rules for \whatsapp and one rule for \signal.
The single rule in \signal was created by merging two rules 
into one, which was necessary
to speed up verification. This transformation is sound,
as we exploit the rule composition result
of~\cite[Appendix~E.2.1]{cheval2022}, validating that
these two rules adhere to their compatibility criterion
\z{canmerge}. The monitored model still observes the implementation-level steps; the merged rule is used to make verification tractable. Previous models also treat this step as atomic.

\subsection{Threat Models}
Our threat model includes Tamarin's predefined network attacker, who can drop, inject, and 
replay any message in the conversation.
Additionally, we strengthen the attacker's capabilities by leaking the setup keys of any user to the network. This is modeled in rules like the following, where the attacker learns the long-term key:
\begin{lstlisting}
rule compromiseLTK:
  [!PrivateIdentityKey(A, ltk_A)] --> [Out(ltk_A)]
\end{lstlisting}

We define three threat models. For the authentication and secrecy properties of Signal and WhatsApp, we define \asignal and \awhatsapp, respectively. For PCS, we use the threat model \apcs.

\paragraph{\asignal} The attacker can compromise any user's setup keys, including the long-term identity key, signed prekeys, one-time prekeys, and one-time post-quantum prekeys. In addition, the attacker can break the Diffie-Hellman assumption. We model this in one additional rule, which allows the attacker to learn the private share corresponding to any honest public Diffie-Hellman key.

\paragraph{\awhatsapp} The attacker can compromise all setup keys of any user, including the long-term identity key, signed prekeys, and one-time prekeys.

\paragraph{\apcs} Following the threat model of the impossibility result for PCS~\cite{DBLP:conf/sp/CremersMN25}, the attacker can compromise the static state of any user. In the \whatsapp and \signal implementations, this translates to the compromise of the long-term identity key.

\subsection{Properties}
We focus on three main properties: authentication, secrecy of the initial root key, and conversation PCS for the end user. The authentication and secrecy properties are comparable to those in the ProVerif analysis of PQXDH~\cite{bhargavanFormalVerificationPQXDH}, while the PCS property is defined as in~\cite{DBLP:conf/sp/CremersMN25}. \Cref{fig:verification-properties} shows the properties for \signal, since it involves a more complicated threat model, and we provide
the corresponding properties for \whatsapp in the artifact.

\begin{figure*}[t]
      \fontsize{8}{9.5}\selectfont
      \setlength{\jot}{0pt}
      \colorlet{bluegray}{MidnightBlue}
      \begin{subfigure}[b]{0.32\textwidth}
            \centering
            \begin{minipage}[t]{\linewidth}
                  $\begin{aligned}
                         & \forall\;\var{\ltka}\;\var{\ltkb}\;\var{\spkb}\;\var{\opkb}\;\var{\kyberb}\;\var{key}\;\tvar{i}.\; \\
                         & \commentRBlue{B completes the key agreement at \tvar{i}}\\
                         &  \fact{RespDone}(\var{\ltkb}, \var{\ltka}, \var{\spkb}, \var{\opkb}, \\
                         & \qquad \var{\kyberb}, \var{key})\,@\,\tvar{i} \\
                         & \commentRBlue{A completed a matching agreement}\\
                         & \Rightarrow (\exists\;\tvar{j}.\; \tvar{j}<\tvar{i} \wedge \\
                         & \fact{InitDone}(\var{\ltka}, \var{\ltkb}, \var{\spkb}, \var{\opkb}, \\
                         & \qquad \var{\kyberb}, \var{key})\,@\,\tvar{j})\\
                         & \commentRBlue{Or A's LTK was compromised before \tvar{i}}\\
                         & {} \mathbin{\|} (\exists\;\tvar{k}.\; \tvar{k}<\tvar{i} \wedge \fact{CompromiseLTK}(\var{\ltka})\,@\,\tvar{k})\\
                          & \commentRBlue{Or B's SPK was compromised before \tvar{i}}\\
                         & {} \mathbin{\|} (\exists\;\tvar{k}.\; \tvar{k}<\tvar{i} \\
                         & \qquad \wedge \fact{CompromiseSPK}(\var{\ltkb}, \var{\spkb})\,@\,\tvar{k})\\
                         & \commentRBlue{Or DH was broken before \tvar{i}}\\
                         & {} \mathbin{\|} (\exists\;\tvar{k}.\; \tvar{k}<\tvar{i} \wedge \fact{BrokenDH}()\,@\,\tvar{k})
                  \end{aligned}$
            \end{minipage}
            \caption{Authentication.}
            \label{fig:verification-property-authentication}
      \end{subfigure}%
      \hfill
      \begin{subfigure}[b]{0.32\textwidth}
            \centering
            \begin{minipage}[t]{\linewidth}
                  $\begin{aligned}
                         & \forall\;\var{\ltka}\;\var{\ltkb}\;\var{\spkb}\;\var{\opkb}\;\var{\kyberb}\;\var{key}\;\tvar{i}\;\tvar{j}.\; \\
                         & \commentRBlue{B completes the key agreement with \var{key}}\\
                         &  \fact{RespDone}(\var{\ltkb}, \var{\ltka}, \var{\spkb}, \var{\opkb}, \\
                         & \qquad \var{\kyberb}, \var{key})\,@\,\tvar{i} \\
                         & \wedge \fact{K}(\var{key})\,@\,\tvar{j}\quad \commentRBlue{and the attacker knows \var{key}}\\
                         & \Rightarrow (\exists\;\tvar{k}.\; \tvar{k}<\tvar{i} \wedge \fact{CompromiseLTK}(\var{\ltka})\,@\,\tvar{k})\\
                         & \commentRBlue{Or B's LTK or SPK was compromised before \tvar{i}}\\
                         & {} \mathbin{\|} (\exists\;\tvar{k}.\; \tvar{k}<\tvar{i} \wedge \fact{CompromiseLTK}(\var{\ltkb})\,@\,\tvar{k})\\
                         & {} \mathbin{\|} (\exists\;\tvar{k}.\; \tvar{k}<\tvar{i} \\
                         & \qquad \wedge \fact{CompromiseSPK}(\var{\ltkb}, \var{\spkb})\,@\,\tvar{k})\\
                         & \commentRBlue{Or both one-time prekeys were compromised}\\
                         & {} \mathbin{\|} (\exists\;\tvar{k_1}\;\tvar{k_2}.\; \fact{CompromiseOPK}(\var{\opkb})\,@\,\tvar{k_1} \\
                         & \wedge \fact{CompromisePQK}(\var{\ltkb}, \var{\kyberb})\,@\,\tvar{k_2})\\
                         & \commentRBlue{Or DH was broken before \tvar{i}}\\
                         & {} \mathbin{\|} (\exists\;\tvar{k}.\; \tvar{k}<\tvar{i} \wedge \fact{BrokenDH}()\,@\,\tvar{k})\\
                         & \commentRBlue{Or later DH break and B's PQ compromise}\\
                         & {} \mathbin{\|} (\exists\;\tvar{k_1}\;\tvar{k_2}.\; \tvar{i}<\tvar{k_1} \wedge \fact{BrokenDH}()\,@\,\tvar{k_1} \\
                         & \wedge \fact{CompromisePQK}(\var{\ltkb}, \var{\kyberb})\,@\,\tvar{k_2})
                  \end{aligned}$
            \end{minipage}
            \caption{Responder secrecy.}
            \label{fig:verification-property-secrecy}
      \end{subfigure}%
      \hfill
      \begin{subfigure}[b]{0.32\textwidth}
            \centering
            \begin{minipage}[t]{\linewidth}
                  $\begin{aligned}
                         & \forall\;\var{\ltka}\;\var{\ltkb}\;\var{key_0}\;\var{key_1}\;\var{key_2}\;\tvar{i_1}\;\tvar{i_2}\;\tvar{j}\;\tvar{t}.\; \\
                         & \tvar{i_1}<\tvar{i_2} \wedge \tvar{i_2}<\tvar{j}\\
                         & \commentRBlue{A and B exchange DH shares to heal}\\
                         &  \wedge \fact{Heal}(\var{\ltka}, \var{\ltkb}, \var{key_0})\,@\,\tvar{i_1} \\
                         &  \wedge \fact{Heal}(\var{\ltkb}, \var{\ltka}, \var{key_1})\,@\,\tvar{i_2} \\
                         & \commentRBlue{No compromise occurs between \tvar{i_1} and \tvar{i_2}}\\
                         & \wedge \neg (\exists\;\var{any}\;\tvar{k}.\; \tvar{i_1}<\tvar{k} \wedge \tvar{k}<\tvar{i_2}\\
                         & \qquad \wedge \fact{Compromise}(\var{any})\,@\,\tvar{k})\\
                          & \commentRBlue{An honest message step follows}\\
                          &  \wedge \fact{StepParty}(\var{\ltka}, \var{\ltkb}, \var{key_2})\,@\,\tvar{j} \\
                         & \wedge \fact{K}(\var{key_2})\,@\,\tvar{t}\quad \commentRBlue{The attacker knows \var{key_2}}\\
                          & \Rightarrow\quad \commentRBlue{A's or B's LTK is compromised after \tvar{i_2}}\\
                          & (\exists\;\tvar{k}.\; \tvar{i_2}<\tvar{k} \wedge \fact{CompromiseLTK}(\var{\ltka})\,@\,\tvar{k})\\
                          & {} \mathbin{\|} (\exists\;\tvar{k}.\; \tvar{i_2}<\tvar{k} \wedge \fact{CompromiseLTK}(\var{\ltkb})\,@\,\tvar{k})
                  \end{aligned}$
            \end{minipage}
            \caption{Post-compromise security.}
            \label{fig:verification-property-pcs}
      \end{subfigure}
      \caption{Tamarin formulations of the main Signal properties considered in our verification.}
      \Description{Three Tamarin lemma formulations for authentication, responder secrecy, and post-compromise security.}
      \label{fig:verification-properties}
\end{figure*}

\paragraph{Authentication} We prove that a responder completing PQXDH has a matching initiator session with the same parameters, unless the attacker has compromised specific keys or already broken the DH assumption (\cref{fig:verification-property-authentication}). The exception for an earlier DH break means that this lemma does not establish authentication against an active quantum attacker.

\paragraph{Secrecy} We show the responder secrecy property in \cref{fig:verification-property-secrecy}. We have also specified and proved the equivalent property for the initiator. These are initial-root-key secrecy properties, not new proofs of DR message-key forward secrecy. Informally,
the initial secret should remain secret except under specific conditions: compromise of the identity key, compromise of prekeys, a break of the DH assumption before the protocol run, or the attacker learning all relevant keys through compromise or a broken primitive. This is also called forward secrecy for the session key~\cite{bhargavanFormalVerificationPQXDH}.
This key-agreement property is distinct from the DR message-key forward secrecy listed separately in \cref{tab:compare-model}.


\paragraph{Post-Compromise Security} We define conversation PCS for the end user, where, informally, any message sent after a healing period should be secure,
as long as the attacker did not compromise the parties again after
that period.
However, this property does not hold in these applications for two reasons.
First, their session layer enables multiple sessions.
Second, once the long-term identity key is compromised, the attacker can start their own sessions in parallel.
The property is falsified in our model as expected (\cref{tab:tamarin-summary}), reproducing the known impossibility result in our session-layer setting.
Modeling a single-session client would sidestep the first issue, but is out of scope here since we target the deployed, session-layer-enabled clients.
We give the falsified property in \cref{fig:verification-property-pcs}.

\subsection{Summary of Results}
The WhatsApp model is constructed from 15 rules modeling the protocol components, and the Signal model from 14. The threat model contributes between 1 and 5 attacker rules, depending on the property under analysis: 3 for \awhatsapp, 5 for \asignal (which adds a one-time post-quantum key compromise rule and a Diffie-Hellman break rule), and 1 for \apcs.
For each model, we verified the authentication, initiator-secrecy, and responder-secrecy lemmas listed in~\cref{tab:tamarin-summary}; reproduced the expected conversation-PCS counterexample from the literature; and proved 12 sanity traces that safeguard the executability of each step of the model. The secrecy lemmas concern the initial root key, while DR message-key forward secrecy is outside our proved properties.
The sanity traces were proved in 4 minutes for \whatsapp and 6 minutes for \signal.
The verification process was run on a Lenovo ThinkPad X1 Carbon Gen 9
with \qty{16}{\giga\byte} of RAM using Tamarin 1.11.0 on the develop branch.
We summarize our results in~\cref{tab:tamarin-summary}.

\begin{table}[t]
   \caption{Tamarin formal analysis summary. Proofs were obtained either automatically using Tamarin's heuristic proof search or by replaying manually constructed proofs. The runtime shows the time needed for Tamarin to find a proof automatically or to verify a stored proof.}
   \label{tab:tamarin-summary}
  \centering
  \begin{tabular*}{\linewidth}{@{\extracolsep{\fill}}p{2.4cm}p{1.9cm}cr}
    \toprule
    \textbf{Property} & \textbf{Threat model} & \textbf{Result} & \textbf{Runtime} \\
    \midrule
    \multicolumn{3}{l}{\textbf{WhatsApp}} & \textbf{\qty[detect-weight=true]{77}{\second}} \\
    \midrule
     			Authentication & \awhatsapp & \cmark & \qty{26}{\second} \\
    Initiator secrecy & \awhatsapp  & \cmark & \qty{2}{\second} \\
    Responder secrecy & \awhatsapp  & \cmark & \qty{30}{\second} \\
                  PCS &  \apcs  & att / \cmark & \qty{19}{\second} \\
    \midrule
    \multicolumn{3}{l}{\textbf{Signal}} & \textbf{\qty[detect-weight=true]{53}{\second}} \\
    \midrule

     			Authentication & \asignal  & \cmark & \qty{13}{\second} \\
    Initiator secrecy  & \asignal  & \cmark & \qty{7}{\second} \\
    Responder secrecy  & \asignal & \cmark & \qty{12}{\second} \\
			PCS &  \apcs  & att / \cmark & \qty{21}{\second} \\
    \bottomrule
  \end{tabular*}
  \begin{flushleft}\footnotesize
   \cmark = verified property
   \quad 
    att / \cmark = expected known attack reproduced (counterexample)
  \end{flushleft}
\end{table}

\section{Related Work}
\label{sec:related-work}

We first survey previous formal analyses of the Signal protocol, then summarize known limitations in the DY model, and finally discuss approaches for aligning formal models with implementations by using fuzzing techniques and static analyses.

\subsection{Formal Analyses of Signal}
\label{sec:formal-analyses}
\begin{table}[t]
\caption{Comparison of Signal models by covered features. Security properties are compared approximately. Prior models target a narrower protocol fragment. Our model covers the full handshake (X3DH+PQXDH), the ratchet, and Sealed Sender. We trade re-proving session/device-level DR forward secrecy for broader coverage and monitorability.}
  \label{tab:compare-model}
  \centering
  \begin{tabularx}{\linewidth}{l@{}c*{3}{>{\centering\arraybackslash}X}}
    \toprule
    \textbf{Feature} & \textbf{\cite{cremers2023a}} & \textbf{\cite{bhargavanFormalVerificationPQXDH}} &\textbf{\cite{linker2025looping}} & \textbf{Ours} \\
    \midrule
    X3DH handshake & \xmark & \cmark & \cmark & \cmark \\
    PQXDH ($\approx$ X3DH + KEM) & \xmark & \cmark & \xmark & \cmark \\
    \midrule
    Ratcheting & \cmark  & \xmark & \cmark & \cmark \\
    \hspace{.5em} Accurate asymmetric ratchet & \xmark  & \xmark & \cmark & \cmark \\
    \hspace{.5em} Accurate symmetric ratchet & \cmark & \xmark & \cmark & \cmark \\
    \hspace{.5em} Message-key derivation & \xmark  & \xmark & \cmark& \cmark \\
    \midrule
    Message formats & \xmark & \xmark & \xmark & \cmark \\
    \midrule
    Authentication & \xmark  & \cmark  & \xmark  & \cmark\\
    Secrecy 	& \xmark   & \cmark & \cmark & \cmark \\
    (DR) Forward secrecy & \cmark$^{s, d}$& \xmark & \xmark & \xmark \\
    Post-compromise security & \cmark$^{s}$ att$^{*, d}$  & \xmark &  \xmark & att$^{d}$ \\
    \bottomrule
  \end{tabularx}
  \begin{flushleft}\footnotesize 
      $(^*)$ = attack found
      \quad
      (s/d) = session / device compromise
  \end{flushleft}
\end{table}


\paragraph{Analysis in the computational model}
The Signal protocol has been analyzed several times in the computational model~\cite{DBLP:conf/eurocrypt/AlwenCD19, cohn-gordonFormalSecurityAnalysis, canetti2022universally, DBLP:conf/crypto/BienstockFGMR22}, including its later-developed post-quantum handshake PQXDH~\cite{hashimoto2025bundled, fiedler2025security, bhargavanFormalVerificationPQXDH}.
\textcite{cohn-gordonFormalSecurityAnalysis} analyze the Double Ratchet protocol in a multistage authenticated key exchange model of two devices without session handling, establishing forward secrecy, post-compromise security, and session-key indistinguishability.
As is customary in the computational model, these works idealize the specification, making direct comparison with the deployed implementation difficult.

\textcite{kobeissi2017} present a TypeScript reimplementation of Signal and derive models from a restricted language subset.
Their implementation omits the symmetric ratchet, supports only a single session at a time, and lacks multi-device support, session replacement, and prekey replenishment.
Our work instead monitors deployed clients and incorporates Signal's Sesame session-handling layer~\cite{signal-doc}, a detailed DR model, and the PQXDH handshake.

\paragraph{Analysis in the DY model} We compare our work with other DY analyses of the Signal protocol, as summarized in~\cref{tab:compare-model}.

\textcite{cremers2023a} introduce the first Tamarin model that includes Sesame.
They show that the DR achieves post-compromise security at the session level, but not at the conversation level due to Signal's multi-session handling (\cref{sec:verification}).
Their model intentionally abstracts from the handshake to retain tractability, using a simplified DR subprotocol and omitting X3DH entirely.
The authors deem a complete model intractable for automated verification in Tamarin, noting that state-of-the-art formal approaches struggle to handle X3DH key agreement and DR accurately even without additional mechanisms.
We extend their model to align with the runtime behavior of Signal Desktop and WhatsApp Web, enabling monitoring (\cref{sec:signal-desktop} and~\cref{sec:whatsapp-web}).
This includes the PQXDH handshake and other details necessary for monitoring,
such as detailed message formats.

\textcite{bhargavanFormalVerificationPQXDH} perform the first formal analysis of the PQXDH protocol,
i.e., the handshake protocol, but not the DR protocol.
They use the verification tools ProVerif and CryptoVerif, the latter
providing guarantees in the computational model.
They analyze authentication and forward secrecy of the session secret under classical and post-quantum threat models. Their guarantees account for the timing of key compromises and cryptanalytic attacks. In particular, their authentication result excludes a DH break before the key exchange~\cite[Theorem~8]{bhargavanFormalVerificationPQXDH}. We prove similar, but coarser, properties for PQXDH in our work, e.g., we do not prove quantum forward secrecy for the initiator
or model cryptanalytic attacks that decapsulate KEM ciphertexts.

\textcite{linker2025looping} model X3DH and DR in Tamarin, capturing both the handshake and ratcheting protocol, but not out-of-order delivery. They prove message secrecy under leakage of setup keys and Diffie-Hellman shares, using a new methodology to reason about the protocol's looping behavior.
In comparison, we prove the secrecy of only the initial secret, instead of the message key. 


\paragraph{Limitations in the DY model}
Tamarin's constraint solver uses unification to determine the possible origins of terms such as encrypted messages or keys. Its built-in Diffie-Hellman theory supports multiplication in the exponent, but omits addition~\cite{tamarin}. \textcite[Chapters~4--6]{jackson2020improving} explains that directly extending unification-based reasoning to the full algebraic structure of DH exponents faces undecidability barriers. He also shows that abstracting away small-subgroup and invalid-curve behavior can hide attacks. Thus, the choice of cryptographic abstraction limits both the protocols that can be represented and the attacks that an analysis can detect. Monitoring alone does not remove these abstraction limits.


The Sesame model~\cite{cremers2023a} abstracts away the X3DH key agreement and focuses on session handling, assuming that key establishment was performed correctly. Our models include the key agreement and ratchet computations needed for monitoring. For verification, we simplify implementation details and merge rules to keep proof search tractable (\cref{sec:verification-models}).

\subsection{Implementation and Fuzzing Studies}
\label{sec:implementation-studies}

Automated model inference has been used to reconstruct finite automata that approximate Signal's implementation from observed I/O and compare them with the specification automaton~\cite{vandam}.
That work finds differences that could prevent future secrecy from holding against attackers with device access.
The approach relies on fuzzing-based model extraction to find attacks, but
reconstructs only a finite-state machine and omits the ratchet state.

Outside of Signal, there are few works that compare protocol models with
existing implementations.
TLSPuffin~\cite{ammann2024} combines fuzzing with a DY attacker to find
implementation-level attacks that can only be triggered deep within the
protocol run. Their work is intended for attack finding, not verification, but
could be used with a monitored implementation to find where it diverges from
the model, thereby helping to improve the implementation or identify missing behaviors in the model.
Tookan~\cite{bortolozzoAttackingFixingPKCS112010} determines the configuration of a PKCS\#11
model from a set of tests run on a real hardware token and translates attacks
found in the model into interactions with it. This methodology is protocol dependent.

Static analyses~\cite{
sprenger2020,
swamy2011,
aizatulin2012, kobeissi2017, nasrabadi2023,
jurjens2008,
bhargavan2010,
polikarpova2012,arquint2022}
can establish conformance without runtime overhead on small examples,
but require substantial expertise
and become increasingly costly as the implementation grows.
The most closely related work, Igloo~\cite{sprenger2020}, soundly links
compositional refinement and separation logic for distributed-system
verification.
\textcite{arquint2022} connect this style of implementation reasoning to
Tamarin models for security protocols.
Both approaches require considerable proof and specification effort.
Verified compilers (cv2ocaml~\cite{cade2015}, cv2fstar~\cite{lipp2022}, and the compiler of~\cite{almeida2013})
do not apply to implementations that are already in use.
In contrast, SpecMon~\cite{specmon} performs runtime monitoring by observing concrete execution traces of the implementation, rather than exploring all possible execution paths symbolically, as static verifiers do. This allows SpecMon to fully support the composition of the X3DH/PQXDH key agreement and the Double Ratchet in one model and to be extended further. Using this capability, we extend the Sesame model to allow for monitoring.

\section{Limitations and Future Work}
\label{sec:limitations}

Although we have demonstrated the applicability of our methodology,
there are still several open problems around functionality, security,
and trustworthiness. Fortunately, many of them motivate interesting
directions for future research.

\subsubsection*{Security: key exfiltration}

Currently, SpecMon provides no protection against \emph{malicious}
applications, as opposed to merely incorrect ones. As we have seen in
\cref{sec:fault-injection}, all protocols that have payloads allow the
exfiltration of secrets. For Signal/WhatsApp, the application simply
encodes the long-term secrets in a chat message to a malicious
outsider. There is no way to detect encodings of secrets, as they are
not known a~priori. This does not contradict the formal
soundness of SpecMon or Tamarin, as the payload (e.g., the content of chat
messages) is considered an atom in the MSR modeling language: the monitor cannot distinguish ordinary text from bytes that encode a long-term key. Thus, the
modeling language and soundness result encode the requirement that payload be
independent of protocol secrets, which, in our setting, is an
assumption about an application we do not fully trust.

As exfiltration cannot be prevented, the secrets must be
hidden from malicious applications. If set up correctly, cryptographic APIs
like PKCS\#11 can (provably~\cite{bortolozzoAttackingFixingPKCS112010, daxHowWrapIt2019})
guarantee key secrecy.
A key challenge here is to find a way to transparently apply this
change, even for applications that do not use PKCS\#11.

\subsubsection*{Functionality: behavior coverage}

We tested our monitoring models through manual GUI interactions and
randomized sequences of UI and external actions (\cref{sec:model-validation}).
There could be other desired implementation behaviors that we missed,
in which case the monitor would reject the trace. For
a closed-source application distributed with the monitor, this
would lead to unexpected termination on the end user's system.
Both UI actions and input from the network can cause such rejections.

Our fuzzer explores combinations of the selected actions but provides no
coverage guarantee. Extending the action set and guiding exploration using
coverage feedback are directions for future work. Existing fuzzing approaches
for UI actions~\cite{memonAutomaticallyRepairingEvent2008} and network
traffic~\cite{ammann2024} could inform these extensions.

\subsubsection*{Deployment: standardized instrumentation}

Our instrumentation is tailored to each application: annotated libraries for
Signal Desktop and DevTools-based runtime wrappers for WhatsApp Web
(\cref{sec:event-extraction}). A standardized instrumentation interface,
provided by vendors, would let monitors like SpecMon attach directly to any
conforming application, supporting a ``don't trust, but verify''
deployment model.
In the browser setting, the monitor could also be deployed in the spirit of
integrity checkers such as Code Verify~\cite{codeverify}.
It could run as a browser extension that observes the application or as a
WebAssembly module alongside the application in the page.

\subsubsection*{Trustworthiness: automated model refinement}

Our current toolchain exposes the gap between verification and monitoring models, but does not automate their reconciliation.
The simplifications that constitute that gap are only informally
justified. This is not unique to our methodology; it only becomes
\emph{visible} here: Tamarin models
frequently omit messages or parts of messages, ignore functionality
believed orthogonal to the argument (e.g., DDoS protections), etc.
Nevertheless, we find that our verification model is much closer
to the implementation than prior models, while remaining
tractable for Tamarin.

Moreover, a
gap between two MSR models
is preferable to a gap between a model and its implementation.
Most importantly, we can reason about this gap. There is work on
formally justifying modeling abstractions specifically on
message terms~\cite{nguyenAbstractionsSecurityProtocol2018}. This does
not fully close the gap: their method is not implemented, and the gap
is not purely about terms. Still, even term-level abstractions already
yield large speedups; mechanized congruence proofs or
automatically simplified models would have wider applications.

\subsubsection*{Scalability: trace rewriting and composable
verification}

Both Signal Desktop and WhatsApp Web communicate with their servers using WebSockets over TLS.
Inside this channel, Signal wraps messages using its Sealed Sender protocol to hide the sender's identity,
while WhatsApp Web establishes the Noise protocol for client--server communication.
We found that SpecMon's trace rewriting mechanism (see
\cref{sec:trace-rewriting}) makes it straightforward to layer
monitoring models to capture nested communication.
Tamarin, by contrast, lacks compositional reasoning, and composition results in comparable tools (e.g., for the applied-pi calculus~\cite{arapinisVerifyingPrivacyTypeProperties2012}) do not match trace rewriting, though channel-specific results~\cite{chevalSecureRefinementsCommunication} could be adapted.

The existing TLS model~\cite{tlstamarin} is very large.
Verifying this model alone already requires about a week of
continuous computation. Analyzing the composed system using the
current holistic analysis methods is prohibitively expensive, 
even though monitoring (which we tested for the intermediate layers,
e.g., Sealed Sender, not for TLS) seems to handle nesting well.

\section{Discussion and Lessons Learned}
\label{sec:discussion}

Our case studies combined verification-oriented and monitoring-oriented modeling in a single methodology.
The lessons below may also apply to other protocols and implementations.

\subsubsection*{One model for verification and monitoring}

Verification and monitoring place different demands on the model.
Verification needs enough abstraction to remain tractable, whereas monitoring needs implementation details such as concrete message formats, helper computations, and persistent state.
These details can substantially increase verification time.
Maintaining a unified model with monitorable and verification variants is therefore harder than performing either task alone, but it provides a shared validation point.
Accepted implementation traces are checked against the monitorable variant, and explicit transformations produce the tractable verification variant (\cref{sec:model-construction,sec:verification-models}).
These transformations are not yet mechanized, so they expose rather than eliminate the remaining abstraction gap.
In our experience, this discipline also produces better-structured models.
Most of the following lessons improved both monitoring performance and verification time.

\subsubsection*{Scope rules by their inputs}

Rule boundaries are marked by input and freshness facts.
We found that a rule should be scoped so that its first observable computation directly uses an input or a freshly sampled value.
This criterion is sufficient for our models and benefited both monitoring and verification.
Future SpecMon versions could automate this check.

\subsubsection*{Avoid accidental nondeterminism}

SpecMon natively supports nondeterministic specifications by exploring all applicable rules, which is flexible but costly.
At the scale of our case studies, it is easy to introduce nondeterminism \emph{accidentally}.
Using SpecMon's debug output, we identified states in which multiple rules were applicable.
We then added differentiating parameters to the relevant facts, which limited exploration to the intended cases.

\subsubsection*{Scope facts for pruning}

The monitor's core work is matching the current configuration against potential rule applications, so its performance depends on how quickly it can prune facts that cannot match.
Pruning on the first argument of a fact is cheaper than pruning on the last.
Scoping facts to a role's identity and local protocol state sped up both monitoring and verification.
Across the tested trace lengths, mean peak process memory increases but remains below \qty{54}{\mebi\byte} for all three workloads (\cref{sec:performance-overhead}).

\subsubsection*{Distinguish protocol messages from modeling artifacts}

Tamarin's \texttt{Out} facts model communication with the adversary, such as publishing public keys.
They are syntactically indistinguishable from protocol messages that the implementation actually sends.
The monitor therefore stored these facts initially but never consumed them because the implementation does not emit corresponding events.
Marking them with the same macro mechanism used for protocol messages allows the monitor to skip storing them.
A model should therefore distinguish messages in the implemented protocol from artifacts of the modeling language.

\subsubsection*{Instrument chokepoints}

A small number of low-level functions can expose a large share of protocol behavior.
In WhatsApp Web, the WebCrypto \texttt{encrypt}, \texttt{decrypt}, and \texttt{sign} functions produced events across X3DH, the Double Ratchet, and the Noise transport layer.
The \texttt{sign} function also covered hashing.
Such chokepoints provide broad coverage with little instrumentation and help locate the remaining protocol-relevant functions in a minified codebase.

\subsubsection*{Reusable methodology}

The instrumentation is reusable.
For other Signal-based applications, whether open or closed source, reusing \libsignal{} instrumentation reduces the manual effort.
Developing the initial WhatsApp Web model and instrumentation took two person-weeks.
Revising the instrumentation, adding fuzzing and out-of-order message support, and rerunning the experiments took one additional person-week.

\section{Conclusion}
\label{sec:conclusion}

\begin{full}
We present the first runtime monitoring of the Signal protocol in production messaging applications.
By applying SpecMon to both Signal Desktop~\cite{signal-d} and WhatsApp Web~\cite{wa}, we demonstrate that runtime monitoring can bridge the verification gap between formal protocol models and real-world implementations.

Our contributions include:
\begin{enumerate*}[label=(\arabic*)]
  \item the most comprehensive monitorable model of Signal to date, combining the X3DH/PQXDH handshake and Double Ratchet protocols with accurate implementation details, including message formats and the Sealed Sender mechanism;
  \item the first formal model of WhatsApp's Signal-protocol variant, revealing previously undocumented implementation differences;
  \item a methodology for monitoring both open-source (Signal Desktop via annotated libraries) and closed-source (WhatsApp Web via browser DevTools) applications; and
  \item empirical evidence that monitoring has low overhead in our measured setting while detecting security-relevant protocol deviations.
\end{enumerate*}

Our evaluation shows successful monitoring of session initialization and symmetric and asymmetric ratcheting. The Signal out-of-order example and the WhatsApp workload containing reordered messages also demonstrate monitoring of out-of-order delivery.
Fault-injection experiments show that SpecMon detects unexpected network outputs, malformed messages, and incorrect uses of cryptographic libraries. Secrets encoded as ordinary message payloads remain outside this detection capability (\cref{sec:limitations}).

For vendors, our work demonstrates a practical path to establish trust in proprietary implementations: provide formal models that serve as executable documentation, integrate runtime monitors to check protocol compliance within the monitored scope, and submit models to rigorous verification---all while maintaining proprietary control over source code.
For researchers, the case studies show that the methodology is viable, flexible, and portable: after the Signal Desktop case study, adapting the workflow to WhatsApp Web was mainly a matter of identifying and instrumenting the relevant implementation functions.
The main remaining opportunity is automation.
Model refinement still requires manual analysis; reducing that effort would make monitorable, verified models easier to maintain as messaging implementations evolve.
\end{full}

\begin{conf}
We present the first runtime monitoring of the Signal protocol in production
messaging applications, applying SpecMon to Signal Desktop~\cite{signal-d} and
WhatsApp Web~\cite{wa}. Our contributions are monitorable models covering the
X3DH/PQXDH handshake, Double Ratchet, and Sealed Sender; the first formal model of
WhatsApp's Signal-protocol variant; and a methodology for monitoring open- and
closed-source applications.

The evaluation demonstrates monitoring of session initialization, symmetric and
asymmetric ratcheting, and out-of-order delivery in both applications, with low
overhead in the measured setting. Fault injection detects unexpected network
outputs, malformed messages, and incorrect uses of cryptographic libraries.
Secrets encoded as ordinary message payloads remain outside this detection
capability (\cref{sec:limitations}).

Automating instrumentation and model refinement would reduce maintenance
effort as applications evolve.
\end{conf}

\section*{Acknowledgments}
\label{sec:acknowledgements}
Moustafa Said completed this work as a master's student at Saarland University, Germany. Kevin Morio and Aurora Naska completed this work as members of the Graduate School of Computer Science, Saarland University, Germany.

This paper was edited for grammar with ChatGPT and Claude.

\printbibliography

\appendix
\crefalias{section}{appendix}

\section{Open Science}
\label{sec:open-science}

\begin{full}
The companion artifact contains our formal models and trace-rewriting rules,
application instrumentation, recorded traces, and the monitor measurements
underlying \cref{fig:performance-processing-time,fig:performance-memory}.
It is available at:
\begin{center}
  \url{https://doi.org/10.5281/zenodo.19892766}
\end{center}
Experiment scripts and a reproduction environment with pinned source revisions
and patches support rerunning verification and recorded-trace monitoring,
including new performance measurements. Plotting scripts regenerate the
processing-time and memory figures from the archived measurements or new runs.
The documentation describes experiment configuration and live monitoring setups.
The latter require application accounts and access to the respective services.
\end{full}

\begin{conf}
The artifact at \url{https://doi.org/10.5281/zenodo.19892766}
includes models, rewriting rules, instrumentation, recorded traces, monitor
measurements, and scripts to verify, replay, rerun measurements, and regenerate
processing-time and memory figures. It pins source revisions and patches and
documents experiment configuration and live setups requiring application
accounts and service access.
\end{conf}

\section{Ethical Considerations}
\label{sec:ethics}

Our WhatsApp experiments were conducted exclusively in the client-side environment of WhatsApp Web using personal test accounts. No unauthorized access to user data or Meta infrastructure occurred. These experiments were conducted for academic purposes and comply with Meta's Bug Bounty Policy guidelines.

We did not identify a concrete bug, security vulnerability, or demonstrable misbehavior in WhatsApp.
Our findings on protocol design and feature adoption in WhatsApp and Signal reflect deliberate engineering choices, despite both using \libsignal{}.

Separately, we found and reported to the developers that
Signal Desktop crashes after sending $4{,}348$ consecutive messages.

\begin{full}
\section{Instrumentation Details}
\label{sec:eval-instrumentation}

\Cref{tab:signal-instrumented-functions,tab:whatsapp-instrumented} list the
instrumented functions for Signal Desktop and WhatsApp Web, respectively.
\end{full}

\begin{conf}
\section{Instrumentation Details}
\label{sec:eval-instrumentation}

\Cref{tab:instrumented-functions} summarizes the instrumentation. The full
version provides extended tables for Signal Desktop and WhatsApp Web~\cite[%
\fullversionlocation{tab:signal-instrumented-functions},
\fullversionlocation{tab:whatsapp-instrumented}]{full}.

\begin{table}[!b]
  \caption{Instrumented functions for Signal Desktop and WhatsApp Web.}
  \label{tab:instrumented-functions}
  \centering
  \begin{tabularx}{\linewidth}{lX}
    \toprule
    \textbf{Function} & \textbf{Symb.\ abstraction} \\
    \midrule

    \multicolumn{2}{l}{\textbf{Signal Desktop}} \\
    \midrule
    \texttt{hmac\_sha256, hkdf}
      & \mbox{$\term{h}(x), \term{hkdf}(\mathit{salt}, \mathit{seed})$} \\

    \texttt{diffie\_hellman}
      & $x^y$ \\

    \texttt{aes\_256\_*\_encrypt/decrypt}
      & \mbox{$\term{senc}(m, k), \term{sdec}(m, k)$} \\

    \begin{tabular}[c]{@{}l@{}}\texttt{handle\_inner\_response}\\ \texttt{send\_request}\end{tabular}
      & \begin{tabular}[c]{@{}l@{}}$\term{recv}(m)$\\ $\term{send}(m)$\end{tabular} \\

    \begin{tabular}[c]{@{}l@{}}\texttt{encapsulate}\\ \texttt{decapsulate}\end{tabular}
      & \begin{tabular}[c]{@{}l@{}}$\aenc(\mathit{ss}, \mathit{pk})$\\ $\adec(\mathit{ct}, \mathit{sk})$\end{tabular} \\

    \texttt{KeyPair.generate}
      & $\term{rand}()$ \\

    \texttt{verify\_signature}
      & $\term{verify}(s, m, \mathit{pk})$ \\

    \midrule

    \multicolumn{2}{l}{\textbf{WhatsApp Web}} \\
    \midrule
    \begin{tabular}[c]{@{}l@{}}\texttt{sign(HMAC,$0^{64}$,y)}\\ \texttt{sign(HMAC,x,y)}\end{tabular}
      & \begin{tabular}[c]{@{}l@{}}$\term{h}(y)$\\ $\term{hkdf}(x, y)$\end{tabular} \\

    \texttt{ecdh(x, y)}
      & $x^y$ \\

    \begin{tabular}[c]{@{}l@{}}\texttt{encrypt(m, k)}\\ \texttt{decrypt(c, k)}\end{tabular}
      & \begin{tabular}[c]{@{}l@{}}$\term{senc}(m, k)$\\ $\term{sdec}(c, k)$\end{tabular} \\

    \texttt{verifyMsgSignalVariant(pk,m,s)}
      & $\term{verify}(s, m, \mathit{pk})$ \\

    \texttt{onmessage(m)}
      & $\term{recv}(m)$ \\

    \texttt{send(m)}
      & $\term{send}(m)$ \\

    \begin{tabular}[c]{@{}l@{}}\texttt{getRandomValues()}\\ \texttt{makeSerializedKeyPair()}\\ \texttt{makeKeyPair()}\end{tabular}
      & $\term{rand}()$ \\

    \bottomrule
  \end{tabularx}
\end{table}

\end{conf}
\begin{full}
\hypersetup{next-anchor=table.caption.53}
\begin{table*}[th]
  \captionsetup{skip=3pt}
  \caption{Instrumented functions for Signal Desktop.}
  \label{tab:signal-instrumented-functions}
  \centering
  \renewcommand{\tabularxcolumn}[1]{m{#1}}
  \begin{tabularx}{\linewidth}{llX}
    \toprule
    \textbf{Function} & \textbf{Symbolic abstraction} & \textbf{Comment} \\
    \midrule
    \texttt{hmac\_sha256} & $\term{h}(x)$ &
      \multirow[c]{2}{=}{Imported from the HKDF module~\cite{hkdf}.} \\
    \texttt{hkdf} & $\term{hkdf}(\mathit{salt}, \mathit{seed})$ & \\
    \midrule
    \texttt{diffie\_hellman} & $x^y$ &
      Curve25519 scalar multiplication~\cite{dalek}. \\
    \midrule
    \texttt{aes\_256\_*\_encrypt} & $\term{senc}(m, k)$ &
      \multirow[c]{2}{=}{AES-256 CBC/CTR operations, partially from RustCrypto~\cite{rustcrypto}.} \\
    \texttt{aes\_256\_*\_decrypt} & $\term{sdec}(m, k)$ & \\
    \midrule
    \texttt{handle\_inner\_response} & $\term{recv}(m)$ &
      \multirow[c]{2}{=}{WebSocket module (\texttt{ws2}) wrappers in \libsignal{}.} \\
    \texttt{send\_request} & $\term{send}(m)$ & \\
    \midrule
    \texttt{encapsulate} & $\aenc(\mathit{ss}, \mathit{pk})$ &
      \multirow[c]{2}{=}{ML-KEM operations provided by \libsignal{} via \texttt{libcrux-ml-kem}~\cite{libsignal,libcrux-mlkem}.} \\
    \texttt{decapsulate} & $\adec(\mathit{ct}, \mathit{sk})$ & \\
    \midrule
    \texttt{KeyPair.generate} & $\term{rand}()$ &
      Randomness generation in \libsignal{}. \\
    \midrule
    \texttt{verify\_signature} & $\term{verify}(s, m, \mathit{pk})$ &
      Signature verification in \libsignal{}. \\
    \bottomrule
  \end{tabularx}
\end{table*}

\hypersetup{next-anchor=table.caption.54}
\begin{table*}[tp]
  \captionsetup{skip=3pt}
  \caption{Instrumented functions for WhatsApp Web.}
  \label{tab:whatsapp-instrumented}
  \centering
  \renewcommand{\tabularxcolumn}[1]{m{#1}}
  \begin{tabularx}{\linewidth}{llX}
    \toprule
    \textbf{Function} & \textbf{Symbolic abstraction} & \textbf{Comment} \\
    \midrule
    \texttt{sign(HMAC,$0^{64}$,y)} & $\term{h}(y)$ &
      \multirow[c]{2}{=}{Imported from the \texttt{crypto.subtle} module~\cite{cryptoapi}.} \\
    \texttt{sign(HMAC,x,y)} & $\term{hkdf}(x, y)$ & \\
    \midrule
    \texttt{ecdh(x, y)} & $x^y$ &
      Base-point multiplication on an elliptic curve, imported from the \texttt{WASignalKeys} module~\cite{wa}. \\
    \midrule
    \texttt{encrypt(m, k)} & $\term{senc}(m, k)$ &
      \multirow[c]{2}{=}{CBC encryption/decryption from \texttt{crypto.subtle}~\cite{cryptoapi}.} \\
    \texttt{decrypt(c, k)} & $\term{sdec}(c, k)$ & \\
    \midrule
    \texttt{verifyMsgSignalVariant(pk, m, s)} & $\term{verify}(s, m, \mathit{pk})$ &
      Signature verification via the \texttt{WASignalSignatures} module~\cite{wa}.
      Arguments are the public identity key, signed prekey, and signature. \\
    \midrule
    \texttt{onmessage(m)} & $\term{recv}(m)$ &
      \multirow[c]{2}{=}{Imported from the WebSocket module~\cite{wa}.} \\
    \texttt{send(m)} & $\term{send}(m)$ & \\
    \midrule
    \texttt{getRandomValues()} &
      \multirow[c]{3}{*}{$\term{rand}()$} &
      \multirow[c]{3}{=}{Randomness generation. Imported from~\cite{cryptoapi} and partially from the \texttt{WASignalKeys} module~\cite{wa}.} \\
    \texttt{makeSerializedKeyPair()} & & \\
    \texttt{makeKeyPair()} & & \\
    \bottomrule
  \end{tabularx}
\end{table*}
\end{full}

\begin{full}
\noindent
We implement 
\ref{it:new} and \ref{it:session-loss}
by removing the common session either from both session tables or,
for \ref{it:session-loss}, only from Parker's.
For \ref{it:send} and \ref{it:receive}, we instrument the sending
functionality; for the latter, we simply run it on Parker.
For
\ref{it:net-swap}
and
\ref{it:net-skip},
we manipulate the network delivery functions by
inserting a proxy into the code that delays the first message for Signal Desktop.
For WhatsApp Web, reordering at the WebSocket proxy layer fails because the
client-server channel is protected by Noise and sequence-number checks reject
reordered frames.
We therefore intercept outgoing messages after Signal-protocol encryption and
before Noise transport encryption, at \texttt{NoiseSocket.sendFrame}.
The instrumentation can hold back a message and release it after later
messages. Each released message passes through the original Noise encryption
function, so the transport sequence numbers remain valid while the enclosed
Signal messages arrive out of order.
Our models support out-of-order delivery.
The artifact includes the instrumentation and fuzzer actions needed to exercise it.
\end{full}

\begin{conf}
Actions \ref{it:new} and \ref{it:session-loss} in
\cref{sec:model-validation} delete the common session from
both parties' tables or only Parker's, respectively.
Actions \ref{it:send} and \ref{it:receive} invoke Monique's and Parker's
instrumented send functions. For \ref{it:net-swap} and \ref{it:net-skip},
a proxy delays Signal Desktop's first message. For WhatsApp Web, we reorder
messages at \texttt{NoiseSocket.\allowbreak sendFrame} between Signal-protocol
and Noise encryption, preserving transport sequence numbers.
\end{conf}

\end{document}